\documentclass[twocolumn,twoside,10pt,journal]{IEEEtran}
\usepackage{epsfig}
\usepackage{subfigure}
\usepackage{amsfonts}
\usepackage{multirow} 
\usepackage{perso}
\usepackage[nolist]{acronym}
\usepackage{graphicx}
\usepackage{cite}
\usepackage{amssymb}
\usepackage{amsmath}
\usepackage{xcolor}
\usepackage{psfrag}
\usepackage{array}
\usepackage{yfonts}
\usepackage[T1]{fontenc}

\IEEEoverridecommandlockouts

\newtheorem{example}{Example}[section]

\newtheorem{theorem}{\indent Theorem}[section]
\newtheorem{assumption}{\indent Assumption}
\newtheorem{corollary}{Corollary}[section]

\newtheorem{lemma}{Lemma}[section]

\newcommand{\D}{\mathrm{d}}

\makeatletter
\newcommand{\thickhline}{%
    \noalign {\ifnum 0=`}\fi \hrule height 1pt
    \futurelet \reserved@a \@xhline
}
\makeatother

\begin{document}
\begin{acronym}
\acro{AWGN}{additive white Gaussian noise} \acro{EXIT}{extrinsic
information transfer} \acro{RA}{random access}
\acro{CCw/oFB}{collision channel without
feedback}\acro{SIC}{successive interference
cancellation}\acro{SN}{slice node}\acro{BN}{burst node}\acro{CSA}{coded
slotted ALOHA}\acro{MAC}{medium access control}\acro{CRA}{coded
random access}\acro{SPC}{single parity-check}\acro{MDS}{maximum
distance separable}\acro{RS}{Reed-Solomon}\acro{SNR}{signal-to-noise
ratio}\acro{D-GLDPC}{doubly-generalized LDPC}\acro{DE}{differential
evolution}\acro{PLR}{packet loss rate}\acro{LDPC}{low-density
parity-check}\acro{DVB}{Digital Video Broadcasting}\acro{RCS}{Return
Channel via Satellite}\acro{BEC}{binary erasure
channel}\acro{MAP}{maximum-a-posteriori}\acro{GJE}{Gauss-Jordan
elimination}\acro{BP}{belief
propagation}\acro{IT}{iterative}\acro{CRDSA}{contention resolution
diversity slotted ALOHA}\acro{IRSA}{irregular repetition slotted
ALOHA}\acro{GA-MAP}{genie-aided maximum-a-posteriori}\acro{p.m.f.}{probability mass function}
\acro{WEF}{weight enumerating function}\acro{GLDPC}{generalized low-density parity-check}
\acro{PAM}{pulse amplitude modulation}
\acro{i.i.d.}{independent and identically distributed}
\acro{MF}{matched filter}
\acrodef{DAMA}{demand assignment multiple access}
\acrodef{DSA}{diversity slotted ALOHA}
\acrodef{CDMA}{code-division multiple access}
\acrodef{VEG}{variance exchange graph}
\acrodef{RFID}{radio frequency identification}
\acrodef{MPR}{multipacket reception}
\acrodef{SIS}{successive interference subtraction}
\acrodef{IIS}{iterative interference subtraction}
\acrodef{QPSK}{Quadrature phase-shift keying}
\end{acronym}

\title{Coded Slotted ALOHA: A Graph-Based Method for Uncoordinated Multiple Access}
\author{Enrico Paolini, {\em Member, IEEE}, Gianluigi Liva,
{\em Senior Member, IEEE},\\ and Marco Chiani, {\em Fellow, IEEE}
\thanks{
The research leading to these results has received funding in part by the Italian Ministry of Education, Universities and Research (MIUR) under Research Projects of Significant National Interest PRIN 2011 ``GRETA'' and in part by the Deutscher Akademischer Austausch Dienst (DAAD) under Fellowship no. 156. The material in this paper was presented in part at the 2011 IEEE International Conference on Communications, the 2011 IEEE Global Telecommunications Conference, the 2012 IEEE First AESS European Conference on Satellite Telecommunications, and the 2012 Asilomar Conference on Signals, Systems, and Computers.}
\thanks{
Enrico Paolini and Marco Chiani are with the Department of Electrical, Electronic, and Information Engineering ``G. Marconi'', University of Bologna, 47521 Cesena (FC), Italy. E-mail: \{e.paolini, marco.chiani\}@unibo.it}
\thanks{
Gianluigi Liva is with Institute of Communication and Navigation of the Deutsches Zentrum fur Luft- und Raumfahrt (DLR), 82234 Wessling, Germany. E-mail: Gianluigi.Liva@dlr.de}
\thanks{Copyright (c) 2014 IEEE. Personal use of this material is permitted.  However, permission to use this material for any other purposes must be obtained from the IEEE by sending a request to pubs-permissions@ieee.org.}
} \maketitle
\date{\today}
\thispagestyle{empty} \setcounter{page}{0}


\begin{abstract}
In this paper, a random access scheme is introduced which relies on the combination of packet erasure correcting codes and successive interference cancellation (SIC). The scheme is named coded slotted ALOHA.
A bipartite graph representation of the SIC process, resembling iterative decoding of generalized low-density parity-check codes over the erasure channel, is exploited to optimize the selection probabilities of the component erasure correcting codes via density evolution analysis. The capacity (in packets per slot) of the scheme is then analyzed in the context of the collision channel without feedback. Moreover, a capacity bound is developed and component code distributions tightly approaching the bound are derived.
\end{abstract}


\markboth
    {}
    {Paolini, Liva, Chiani: Coded Slotted ALOHA} %

{\pagestyle{plain} \pagenumbering{arabic}}


\section{Introduction}\label{sec:Intro}

\IEEEPARstart{R}{andom} multiple access has traditionally represented a popular solution
for wireless networks. The slotted ALOHA protocol \cite{Abramson:ALOHA,Roberts72:ALOHA,abramson94:multiple,Gallager:DataNetworks},
for example, is still employed for the initial access in both cellular terrestrial and satellite
communication networks \cite{Morlet07_RCS}. 
As opposed to \ac{DAMA} protocols,
random access schemes let a common channel to be dynamically and opportunistically shared by a 
population of users, among whom only a low level of coordination (or even no coordination at all) is permitted.
In practice, the impossibility to establish a sufficient level of coordination among the users wishing to access the channel may be due to several reasons, 
for instance, to a lack of global information, to intolerable delays introduced by coordination establishment, to a too large user population size, 
or to the sporadic and unpredictable nature of users' access activity. As a result of the uncoordinated users' transmissions packets may experience 
\emph{collisions}, traditionally requiring the retransmission of (some of) the involved packets, resulting in stability issues. 
For a random multiple access system in which each user is equipped with a buffer of infinite size to store packets that have not yet been transmitted or 
correctly received, stability is often intended as the property that all users' queues admit a limiting distribution (a formal definition may be found, for instance, 
in \cite{szpankowski94:stability}). The \emph{stability region} of random multiple access systems under different interacting queue 
settings has been deeply investigated in several works, such as  \cite{eph:98,eph:98b,eph:06,eph13:cognitive}.

A new light on random access techniques has recently been cast 
by the observation that iterative signal processing can largely improve the transmission efficiency, 
rendering the throughput achievable by random access schemes competitive with that typical of coordinated protocols. 
In this respect, \ac{SIC} techniques turned out to represent a major breakthrough, enabling collisions to be favorably exploited 
instead of being regarded simply as a waste. These
techniques share the feature of cancelling the interference caused by
collided packets in the slots where they have been transmitted
whenever a clean (i.e., uncollided) copy of them is detected.
These advances have opened a completely new perspective in uncoordinated protocols, paving the way to dramatic performance improvements.

The \ac{CRDSA} scheme proposed in \cite{DeGaudenzi07:CRDSA}, for example, exploits \ac{SIC} in the framework of satellite access networks to remarkably improve the performance of 
the \ac{DSA} technique \cite{Rappaport83:DSA}, consisting of transmitting each packet twice
over a \ac{MAC} frame. Almost contemporaneously to \cite{DeGaudenzi07:CRDSA}, interference cancellation was employed within the SICTA protocol \cite{Giannakis07:SICTA} 
and, slightly later, within the ZigZag protocol \cite{Katabi2008:ZigZag}. The SICTA protocol exhibits conspicuous performance gains over collision resolution algorithms
working on trees \cite{massey81:collision}. The ZigZag technique, combining packet repetitions and random packet jitters,
was proposed as an effective countermeasure to collisions due to the hidden terminal problem in wireless local area networks. 
More recently, \ac{IRSA} was introduced in \cite{Liva11:IRSA} to provide a further
throughput gain over \ac{CRDSA}, by allowing
a variable and judiciously designed repetition rate for each packet. (The \ac{IRSA} scheme may be regarded a special case of
the access technique proposed in this paper, as it will be explained later.) Moreover, an improvement to the original 
ZigZag approach, exploiting soft message-passing and named SigSag, was presented in \cite{Dimakis2011:SigSag}. Both 
\cite{Liva11:IRSA} and \cite{Dimakis2011:SigSag} identified a key connection between the \ac{SIC} process and iterative 
message-passing algorithms on sparse graphs. This connection was exploited in \cite{Liva11:IRSA} to design via density evolution
\cite{studio3:richardson01capacity} \ac{IRSA} configurations with remarkably high peak throughput values, and in \cite{Dimakis2011:SigSag} to 
interpret the original ZigZag algorithm as an instance of the sum-product algorithm on factor graphs \cite{Kschischang2001:factor} and, consequently, to
develop a soft version of it. Additional significant works in the area are \cite{Popovski2012:comlet,Popovski2013:rateless} in which a ``frameless'' version of \ac{IRSA} has been
proposed in analogy with rateless codes, \cite{Pfister2012:IRSA} in which an \ac{IRSA} configuration based on the ``soliton distribution''
has been developed achieving a throughput equal to $1\,\mathrm{[packets/slot]}$, and \cite{Kissling11:ICC} in which an unslotted version of \ac{CRDSA} and \ac{IRSA} 
was investigated. It is worth observing that \ac{SIC} techniques have also been successfully exploited to enhance access protocols 
beyond random access. An iterative receiver for asynchronous
\ac{CDMA} systems, exploiting interference cancellation, was for example proposed in \cite{Schlegel2006:CDMA}. Based on an exchange 
of extrinsic information (in a turbo-like fashion) between the interference canceller and the error correcting decoders of individual users, the decoder exhibits very good performances upon a careful design.

While in random access systems communication reliability is typically achieved via retransmissions, 
the problem of recovering from collisions may also be tackled from a different perspective, i.e., from a forward error correction viewpoint. 
A fundamental work in this research area is \cite{Massey85:collision_channel}, in which the capacity region of a ``collision channel without feedback'' (i.e., a multiple access channel on which collisions are unavoidable while reliability cannot be ensured by retransmissions due to the lack of a feedback channel to notify successful transmissions or collisions) was analyzed
and a coding scheme achieving capacity over such channel 
was developed. In the setting considered in \cite{Massey85:collision_channel} collisions are caused by asynchronous (either slot-aligned or unslotted) users' transmissions 
and the multiaccess communication strategy is based on  erasure correcting codes %
and on assigning different periodic protocol sequences to different users, each sequence specifying the slots in which the corresponding user is allowed to access the channel. 
In this way, a symmetric capacity\footnote{This is  the maximum sum-rate for a point in the capacity region under the hypothesis that all users have the same information rate.}
equal to $1/e\,\mathrm{[packets/slot]}$ is achieved as the number of users accessing the channel tends to infinity, both in the slotted and in the unslotted case. 
Although simple and effective, the approach in
\cite{Massey85:collision_channel} poses some coordination challenges, especially
for a large (and varying) number of users, since user protocol sequences must be jointly assigned
\cite{Hui84:CCwoFB,Thomas00:Capacity_Wireless_Collision}. Subsequent works elaborated on the system considered in \cite{Massey85:collision_channel}. 
In \cite{Grant05:CCwithRecovery} the capacity region in the slot-synchronized case was analyzed, under
the more general setting in which collision are not fully destructive due, for instance, to the adoption
of multiuser detection techniques \cite{SIC_Verdu}. Moreover, in \cite{Shum09:Shift_Invariant_Protocol} several 
properties of shift-invariant protocol sequences (ensuring a constant throughput to each user
regardless transmissions offsets) were exposed, along with design strategies for such sequences.

In this paper an extension of the \ac{IRSA} access strategy proposed in \cite{Liva11:IRSA}, dubbed
\ac{CSA}, is proposed. As opposed to \ac{IRSA} and to the above-reviewed schemes exploiting \ac{SIC} in the framework of random access, 
in the new scheme user packets are \emph{encoded} prior to transmission in the \ac{MAC} frame, instead of being simply \emph{repeated}. 
The encoding operation is performed through local component codes (all having the same dimension) randomly drawn by the users, in an uncoordinated fashion, 
from a set of component codes. This latter set together with the \ac{p.m.f.} according to which users pick their codes represent the design parameters
of the proposed access scheme. On the receiver side, \ac{SIC} is combined with decoding of the local component codes to recover from collisions. 
Exploiting a bipartite graph representation, density evolution equations for \ac{CSA} on the collision channel are derived, allowing the analysis of the \ac{SIC} process in an asymptotic setting and leading to the definition of the ``capacity'' of the scheme in a retransmission-free context. It is proved that the scheme is asymptotically reliable on the collision 
channel even without retransmissions. More specifically, in the limit where the \ac{MAC} frame length
and the user population size both tend to infinity (their ratio remaining constant), a vanishing packet loss probability  is guaranteed for channel loads not greater than the asymptotic throughput.

The \ac{IRSA} access scheme can be seen as an instance of \ac{CSA}, where all local component codes are repetition codes. For this reason, \ac{CSA} retains all advantages of \ac{IRSA} in terms of uncoordinated access, equal medium access opportunities for all users, and low complexity processing performed by the users, while overcoming the main weakness of the \ac{IRSA} protocol. As discussed in the next section upon addressing the system model, in fact, while the maximum \emph{rate}\footnote{The rate of the access scheme is formally defined in Section~\ref{subsection:encoding_decoding}.} for an \ac{IRSA} scheme able to reliably operate without retransmissions (up to some value of the load) is $1/2$, a reliable \ac{CSA} scheme can be designed for any rate between $0$ and $1$. If, on the one hand, replacing repetition codes with generic linear block codes may appear as the simplest generalization which allows to overcome the \ac{IRSA} limitation in terms of supportable rates, on the other hand this change defines a framework sufficiently general to include, as marginal variations, several other related access schemes that may be obtained by relaxing some of the conditions in the \ac{IRSA} paradigm. Among them, the introduction of mild forms of coordination among the users aimed at improving the throughput, the introduction of mechanisms for making the traffic generated by some users priority with respect to the traffic generated by the other users, and the introduction of forms of inter-frame processing to resolve the collisions.

With respect to \ac{IRSA}, the \ac{CSA} access protocol is particularly useful in those contexts in which efficiency in terms of transmitted energy is required. In fact, the transmitted energy per packet required by \ac{CSA} is higher than that required by pure slotted ALOHA by a factor that is equal to the ratio of the expected length of the component code drawn by the generic user to the (common) dimension of the component codes, i.e., a factor equal to the inverse of the rate of the scheme. Therefore, the use of local codes with low rates, as it is the case for the repetition codes used in \ac{IRSA}, results in low energy efficiencies, as further discussed in Section~\ref{subsection:encoding_decoding}. Conversely, \ac{CSA} is able to overcome this limitation by admitting high-rate erasure codes as local component codes, allowing in principle rates of the access scheme arbitrarily close to $1$ and thus extending the trade-off between energy efficiency and sustainable channel traffic.

In the high rate regime, i.e., for rates of the access scheme larger than $1/2$, the \ac{CSA} protocol must rely on component codes with large enough dimension and high enough coding rate. In the low rate regime, i.e., for rates of the access scheme lower than $1/2$, any \ac{CSA} scheme has an \ac{IRSA} counterpart. Although advantages of \ac{CSA} protocols over \ac{IRSA} ones in terms of peak throughput transpire from our results over the whole range of low rates, these advantages are so small at very low rates (e.g., rates less than $1/3$) that \ac{IRSA} protocols should be preferred in this rate region due to their design and to operational simplicity. On the contrary, when \ac{IRSA} is operated at rate $1/2$ or close to it, the lack of freedom in the definition of the probability with which a component repetition code is selected results in visible performance degradation \cite{Liva11:IRSA}. Owing to its flexibility in selecting high-rate local codes, \ac{CSA} can rely on a broader set of component codes, allowing a careful definition of the probability with which each of them is picked at any target rate. As a result, for rates comprised between $1/3$ and $1/2$, \ac{CSA} outperforms \ac{IRSA} remarkably, even using very simple binary two-dimensional component codes.

References \cite{paolini10:random,helleseth97:information,Ashikhmin:AreaTheorem,paolini09:stability} are particularly relevant to the present work. Elaborating on some of the results developed in these works, a complete characterization and a systematic design methodology of \ac{CSA} access schemes is presented and a framework for the analysis of CSA-related schemes is defined. In the process, an upper bound on the sustainable traffic for a given rate of the scheme is developed elaborating on the Area Theorem in the context of coding for erasure channels. It is illustrated in the numerical result section how, moving from CSA to IRSA, it is possible to perform closer to the bound, as previously mentioned, and that this performance advantage tends to become more evident as the dimension of the component codes increases. Moreover, an interpretation of the performance of various CSA schemes as a trade-off between energy efficiency and sustainable traffic is proposed.
 
In terms of possible applications, the capability of \ac{CSA} to guarantee communication reliability even without
retransmissions makes it an interesting opportunity for multiple access problems characterized by a potentially very large population of users
(in which case the level of coordination required by \ac{DAMA} protocols cannot be achieved)
and in which the use of retransmissions poses some problems. 
Examples of such applications are wireless sensor networks with a high density of sensor nodes or 
\ac{RFID} systems with a high density of tags. Satellite networks are also a potential application.

This paper is organized as follows. The \ac{CSA} encoding and decoding procedures, along with the adopted notation,
are introduced in Section~\ref{sec:System_Model}. An asymptotic analysis of
the \ac{CSA} decoding process, based on an analogy with iterative decoding of modern codes on graphs, is 
presented in Section~\ref{sec:CSA_Bipartite}, while in Section~\ref{sec:Capacity} an upper bound on the capacity
of the scheme (to be defined later) is developed. Numerical results are presented in Section~\ref{sec:Results} to illustrate
the effectiveness of the asymptotic analysis in designing \ac{CSA}
configurations for a finite number of users. Conclusions follow in Section~\ref{sec:Conclusions}. Results supporting 
some assumptions made during the analysis and an alternative proof of the bound in Section~\ref{sec:Capacity}
are presented in the appendices.


\section{\ac{CSA} System Model}\label{sec:System_Model}


\subsection{Preliminaries}\label{subsec:preliminaries}

We consider a slotted random access scheme where slots are
grouped in \ac{MAC} frames, all with the same length $M$ (in slots).
Each slot has a time duration $T_{\mathsf{slot}}$, whereas the \ac{MAC} frame
is of time duration $T_{\mathsf{frame}}$, so $M=T_{\mathsf{frame}}/T_{\mathsf{slot}}$. 
The total number of users in the system is $N = \alpha M$, where $\alpha$ is the normalized user population size.\footnote{Even if this is not mathematically necessary for the technical results presented in the following, the population size should be thought as large with respect to the number of available slots per frame, i.e., $\alpha \gg 1$. Moreover, it is useful (even if, again, not strictly necessary) to think of users characterized by
a \emph{sporadic} activity, i.e., characterized by an activation probability (defined later) $\pi \ll 1$. This justifies the use of random access schemes instead of \ac{DAMA} ones.} 
Each user is frame- and slot-synchronous 
and attempts at most one \emph{burst} (i.e., packet) transmission per \ac{MAC} frame. Neglecting guard
times, the time duration of a burst is $T_{\mathsf{slot}}$. 

At the beginning of a \ac{MAC}
frame each user generates a burst to be transmitted within the
frame with probability $\pi$, where $\pi$
is called the \emph{activation probability}. Users attempting the
transmission within {a} \ac{MAC} frame are referred to as the
\emph{active} users for that frame. Since each user becomes active independently of the other users, 
the number of active users for a frame is modeled by a random variable $N_{\mathsf a}$ (the subscript ``$\mathsf a$'' reminding the word ``active'')
which is binomially distributed with mean value
$\mathbb{E}[N_{\mathsf a}]=\pi N$. The instantaneous channel load is
\begin{equation}\label{eq:instantaneous_G}
G_{\mathsf a}=\frac{N_{\mathsf a}}{M}
\end{equation}
while the expected channel load 
(representing the expected number of burst transmissions per slot) is
\begin{align}\label{eq:average_G}
G =\frac{\mathbb{E}[N_{\mathsf{a}}]}{M}=\pi \alpha\, .
\end{align}
Clearly, for constant normalized population size $\alpha$ we have $G_{\mathsf a}=G + o(1)$ as $M \rightarrow \infty$.


\subsection{Encoding and Decoding Procedures}\label{subsection:encoding_decoding}

The proposed access scheme works as follows. Prior to
transmission, the burst of
an active user is divided into $k$ information (or data) \emph{segments}, all of the same
length in bits. The $k$ segments are then encoded by the user via a packet-oriented linear
block code generating $n_h$ encoded segments, all of the same length as the data segments. 
For each transmission, the $(n_h,k)$ code is chosen randomly by
the user from a set $\mathcal{C}=\{\mathscr{C}_1,\mathscr{C}_2,\dots,\mathscr{C}_{\theta}\}$ of $\theta$ 
\emph{component codes}. Note that the set $\mathcal C$ is known also to the receiver. 
Unless explicitly stated, all component codes will be assumed to be binary. For $h\in\{1,2,\dots,\theta\}$ the code $\mathscr{C}_h$ has
length $n_h$, dimension $k$, rate $R_h=k/n_h$, and minimum distance $d_h \geq 2$.
Moreover, it has no idle symbols. At any transmission, each user draws its local code from the set $\mathcal{C}$ independently of all its previous choices and
without any coordination with the other users. The code is picked according to a
probability mass function (p.m.f.) {\boldmath
$\Lambda$}$=\{\Lambda_h\}_{h=1}^{\theta}$ which is the same for all
users. A user adopting code $\mathscr C_h$ for transmission in a \ac{MAC} frame is referred to as a type-$h$ user for that frame
and an encoded segment associated with the user
as a type-$h$ segment. For $h \in \{1,\dots,\theta \}$, a type-$h$ encoded segment is equipped with information about the
user it is associated with and about the component code picked by the user. 
Moreover, it is equipped with pointers to the other $n_h-1$ 
encoded segments.\footnote{In practical implementations, the overhead due to
the inclusion of pointers in the segment header may be reduced by adopting more efficient
techniques. For fixed $k$, one may include in the segment header the code index $h$ together with a
random seed, out of which it is possible to reconstruct (by a pre-defined pseudo-random number
generator) the positions of the $n_h$ segments.} 

Encoded segments are further encoded via a physical layer
code before transmission over the multiple access channel. The time duration of each 
transmitted segment is $T_{\mathsf{segment}}=T_{\mathsf{slot}}/k$.
Every slot in the \ac{MAC} frame is divided into $k$ \emph{slices}, each of the same time duration $T_{\mathsf{segment}}$ as encoded segments. Hence, up to $k$ segments may be accommodated
in the same slot and the \ac{MAC} frame may be thought as composed of $kM$
slices.\footnote{
The definition of \ac{MAC} frame as sequence
of $M$ slots is instrumental to the definition of instantaneous and expected loads $G_{\mathsf a}$
and $G$ only. The actual minimum units that can be allocated to a segment transmission
are the slices.} The $n_h$ segments are transmitted
by a type-$h$ active user over $n_h$ slices picked uniformly at random. 
We define the \emph{rate} of the scheme as 
\begin{equation}\label{eq:rate}
R=\frac{k}{\bar{n}}
\end{equation}
where 
\begin{equation}\label{eq:n_average}
\bar{n} = \sum_{h=1}^{\theta} {\Lambda_h} n_h
\end{equation}
is the expected length of the code picked in $\mathcal C$. 
Note that $\Delta E = 10 \log_{10} (\bar{n}/k) = - 10 \log_{10} R$ represents the increment (in dB) of energy per burst with respect to pure SA without retransmissions. 
Note also that, if all component codes in the set $\mathcal C$ are repetition codes ($k=1$), then the \ac{IRSA} scheme is obtained as a special case of \ac{CSA}.
While only rates $0 < R \leq 1/2$ can be obtained with \ac{IRSA}, the \ac{CSA} scheme is more flexible in that all rates $0 < R < 1$ are in principle possible. In particular, to obtain a \ac{CSA} scheme of rate $R$ the minimum dimension $k$ of the component codes is given by $\lceil R / (1-R) \rceil$.

\begin{figure}[t]
\begin{center}
\includegraphics[width=1.0\columnwidth,draft=false]{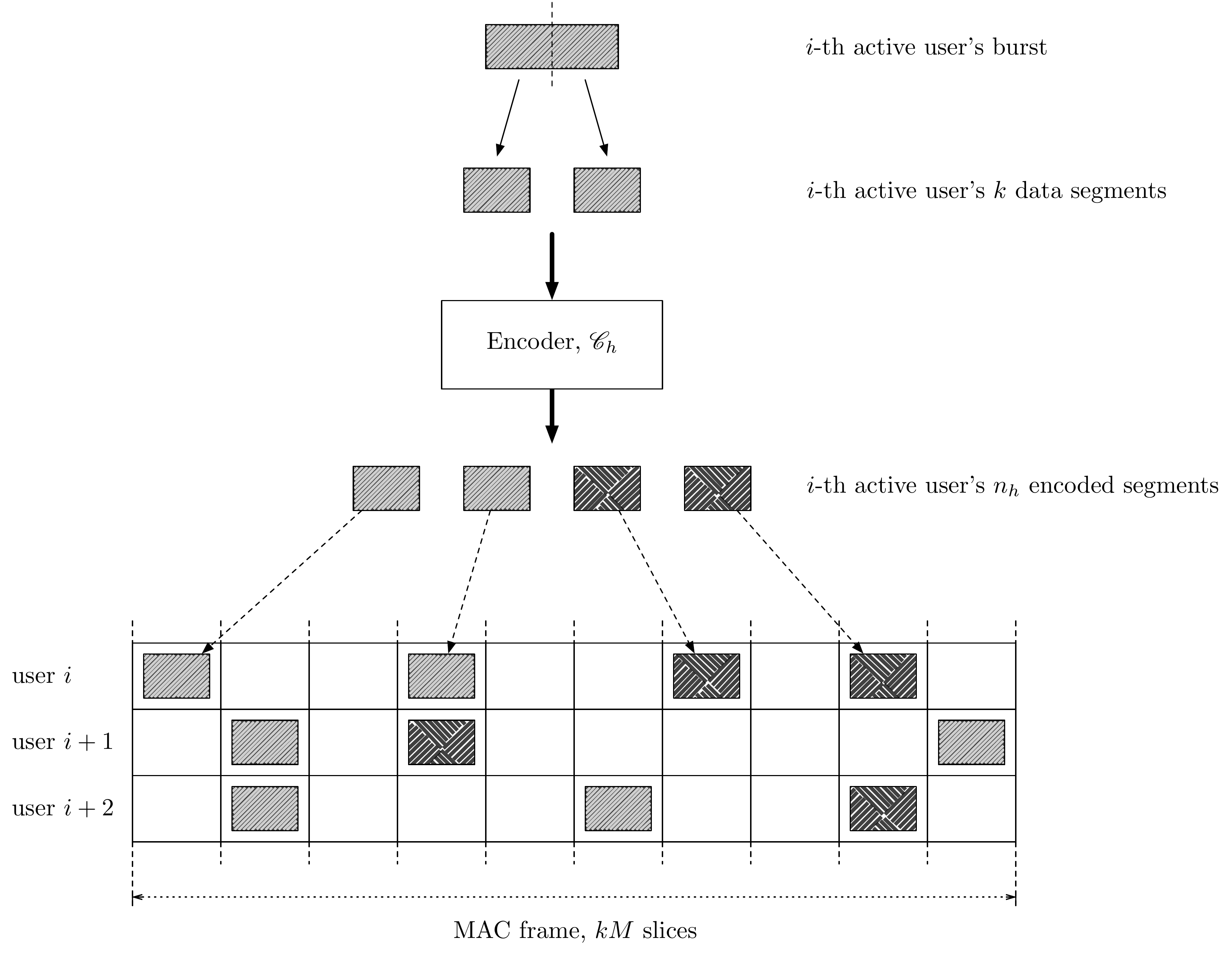} 
\end{center}
\caption{Model of the \ac{CSA} access scheme. Each user being active at the beginning of a \ac{MAC} frame splits his bursts into $k=2$ data segments. User $i$
encodes his data segments via a $(4,2)$ linear block systematic encoder, while users $j$ and $l$ through $(3,2)$ linear block systematic encoders. The darkened rectangles represent parity segments generated by the encoders.}
\label{fig:CRA_Model}
\end{figure}

\medskip
\begin{example}\label{example:encoding}
In Fig.~\ref{fig:CRA_Model} a pictorial representation of the encoding and transmission 
process is provided for the case of $N_{\mathsf a}=3$ active users (indexed as user $i$, user $j$, and user $l$) and $k M=10$ slices 
(indexed from $1$ to $10$). Each burst is split into $k=2$ information segments.
Out of the three users, user $i$ employs a $(4,2)$ linear block code (code $\mathscr C_h \in \mathcal C$) while user $j$ and user $l$ 
employ $(3,2)$ linear block codes. User $i$ performs systematic encoding of its two data
segments, generating two parity segments. The four encoded segments are then transmitted 
into the \ac{MAC} frame slices of indexes $1$, $4$, $7$, $9$. The encoded segments 
of users $j$ and $l$ (performing systematic encoding as well) are transmitted in slices
of indexes $2$, $4$, $10$ and $2$, $6$, $9$, respectively. In the example physical layer 
coding is not represented.
\end{example}

\medskip
In Example~\ref{example:encoding} all users perform systematic encoding of their data segments. Indeed, as it will become clear in Section~\ref{sec:CSA_Bipartite},
the performance of the proposed access scheme does not depend on the specific choice of the generator matrix for each code $\mathscr C_h \in \mathcal C$,
so that a systematic segment encoding process may always been assumed. 

On the receiver side decoding is performed as follows. Segments that are received in clean slices 
(i.e., segments not experiencing collisions) are first decoded at physical layer and information about 
the relevant user, the code $\mathscr C_h \in \mathcal C$ adopted by the user, and the positions
of the other $n_h-1$ segments in the \ac{MAC} frame are extracted. For each active user the receiver
becomes aware of, \ac{MAP} erasure decoding of the code $\mathscr C_h \in \mathcal C$ adopted 
by the user is performed in order to recover as many encoded segments as possible for the user. 
Recovered segments may now be exploited in order to subtract their contribution of interference 
in those slices where collisions occurred. This procedure combining \ac{MAP} erasure decoding 
of the codes employed by active users to encode their data segments and \ac{SIC} is iterated 
until either all slices have been cleaned (and then all bursts have been successfully 
decoded) or collisions persist but no further encoded segments can be recovered 
via \ac{MAP} erasure decoding. Note that the receiver is not a priori aware of the number 
of users becoming active and transmitting a burst in the current \ac{MAC} frame. Note also
that we have implicitly assumed that the receiver is always able to discriminate between 
``empty'' segments and segments where users' waveforms have been received, and 
that collisions are always detected by the receiver, 
even if information neither about the number of users causing the collision 
nor about the single colliding segments can be extracted from the waveform received in the 
corresponding slice. This is reasonable,
for example, when the segment header comprises an integrity control field that is checked,
on the decoder side, after physical layer decoding. 

\medskip
\begin{example}\label{example:decoding}
With reference again to Fig.~\ref{fig:CRA_Model}, assume that all users employ binary 
linear block codes. Specifically, assume that user $i$ encodes its two 
data segments via a $(4,2)$ code with generator matrix $\mathbf G = [1011,0110]$, and that
both user $j$ and user $l$ employ a $(3,2)$ \ac{SPC} code. A collision is detected by the 
receiver on slices with indexes $2$, $4$, and $9$, while interference-free segments are received on
slices with indexes $1$, $6$, $7$, and $10$. It is easy to recognize that \ac{MAP} erasure decoding
of the block code employed by user $i$ allows to recover the two missing segments of this user.
The contributions of interference of these two segments can then be subtracted from the corresponding slices (of indexes $4$ and $9$), cleaning the segments transmitted by user $j$
in slice $4$ and by user $l$ in slice $9$, respectively. Iterating the process, 
\ac{MAP} erasure decoding of the \ac{SPC} codes employed by user $j$ and user $l$ allows
to recover all of the segments transmitted by the two users.
\end{example}


\subsection{Channel Model}\label{subsection:channel_model}

When a segment of some user is recovered via \ac{MAP} erasure decoding, 
a correct implementation of interference cancellation (i.e., cancellation of the 
contribution of interference of this segment in the
corresponding slice) imposes the estimation of channel parameters such as 
the delay, the frequency offset, and the phase offset. Algorithms to efficiently perform 
this estimation have been discussed in \cite{DeGaudenzi07:CRDSA,Liva11:IRSA}. 
Nonetheless, throughout the paper we will adopt 
a channel model in which ideal interference cancellation is assumed. As 
discussed in Section~\ref{sec:CSA_Bipartite}, this model has the advantage to establish a direct connection between the proposed random access scheme and
iterative erasure decoding of a generalization of \ac{LDPC} codes. This bridge 
enables both a simple analysis of the \ac{SIC} process, leading to 
the definition of key performance parameters such as the asymptotic threshold, 
and a simple yet effective access scheme design, in terms 
of selection of the component codes in $\mathcal C$ and of their \ac{p.m.f.} \mbox{\boldmath $\Lambda$}. 

In each slice of the \ac{MAC} frame
the decoder may detect a ``silence'' (no active user has transmitted in that segment), a signal
corresponding to a unique segment, or a signal 
being the result of a collision. As discussed in the previous subsection, 
it is assumed that the decoder can always discriminate between these three events: In case a collision is detected, the observed signal 
provides no information to the decoder about the number and the values of colliding segments.\footnote{This is typical of ``collision channel'' models. A more general setting (not addressed in this paper) is represented by a standard \ac{MPR} channel model \cite{eph:06}, 
in which a packet has a certain probability of being correctly received even in presence of interference from other packets transmitted in the same slot.} Moreover, segments 
not experiencing collisions are always correctly received. This is reasonable when 
a good physical layer channel code is used to individually encode each segment and when the \ac{SNR} 
on the link is sufficiently high. We may better summarize our simplifying assumptions as follows.

\begin{assumption}\label{assumption:1}
Collisions are always detected by the receiver.
\end{assumption} 

\begin{assumption}\label{assumption:2}
All users are within the range of detectability and decodability of the receiver.
\end{assumption}
\begin{assumption}\label{assumption:3}
Interference cancellation is ideal, as so is the estimation of the channel parameters necessary
to perform it.
\end{assumption}

Due to Assumption~\ref{assumption:2} when a segment experiences no collisions it is always correctly detected and decoded, and it is useful for the purposes of interference cancellation process. Moreover, when a segment is involved in a collision with other $d-1$ segments and the interference cancellation algorithm is able to cancel the contribution of interference of these $d-1$ segments, the recovered segment is correctly detected and decoded and, again, it becomes useful for the purposes of interference cancellation process.\footnote{A slightly more general channel model may be obtained by relaxing Assumption~\ref{assumption:2}, as it is done in \cite{ivanov15:error}.} Moreover, due to Assumption~\ref{assumption:3}, hereafter we will use the terminology \emph{interference subtraction} instead of interference cancellation, as it suggests perfect removal of a contribution of interference.


\section{Bipartite Graph Model and Density Evolution Analysis}\label{sec:CSA_Bipartite}

\begin{figure*}[!t]
\begin{center}
\includegraphics[width=1.3\columnwidth,draft=false]{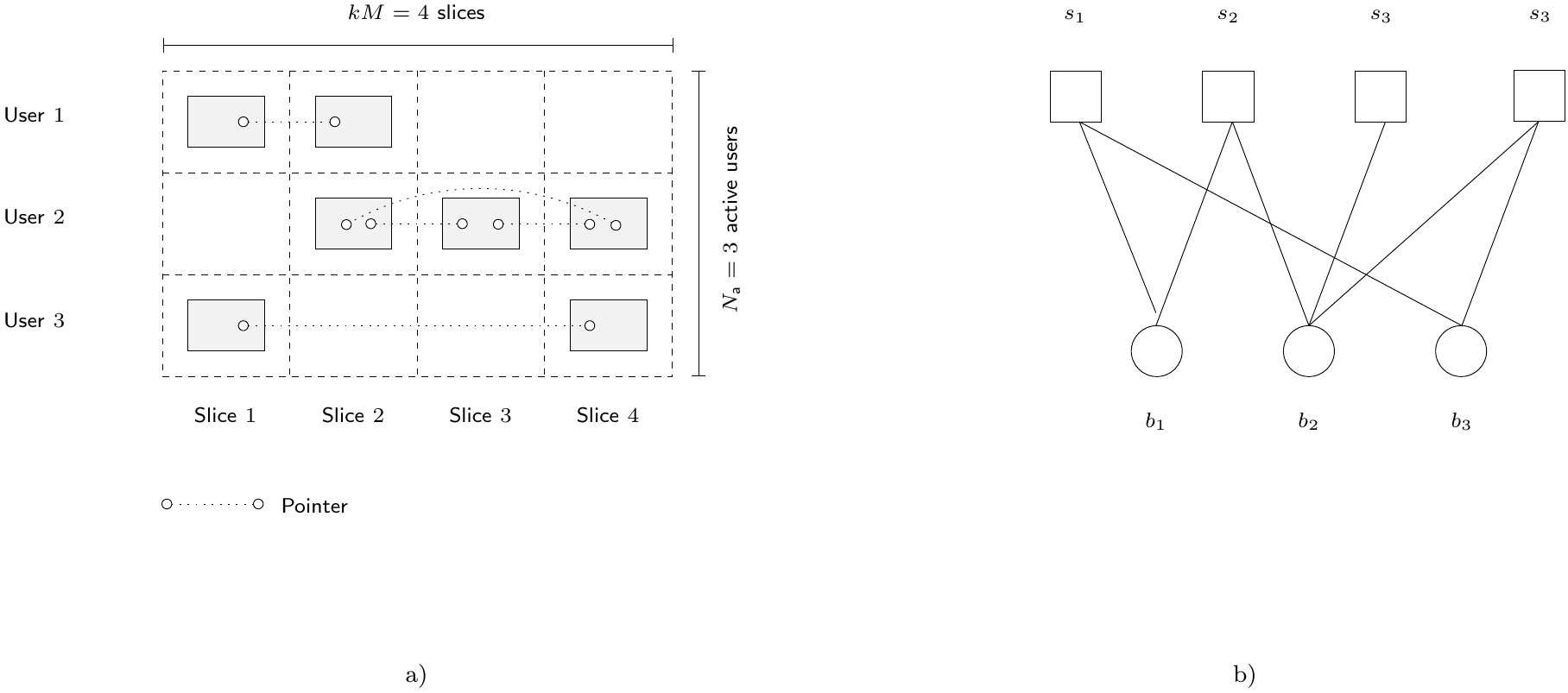}
\end{center}
\caption{On the left, a \ac{MAC} frame made by $kM=4$ slices, with $N_{\mathsf a}=3$ active users. Each user is employing a repetition code. On the right, the bipartite graph representation is provided.}\label{fig:example_TIT}
\end{figure*}

Considering an instantaneous population of $N_{\textsf a}$ active users and a \ac{MAC} frame of $M$ slots, 
the frame status can be described by a bipartite graph, $\msr{G}=(B,S,E)$, consisting of a set $B$ of $N_{\textsf a}$
\textit{burst nodes} (one for each active user), a set $S$ of $M$
\textit{slice} \textit{nodes} (one for each slice in the frame), and a
set $E$ of edges. An edge connects a \ac{BN}
 $b_i\in B$ to a \ac{SN} $s_j\in S$ if and only if the $j$-th slice has been selected by the
$i$-th active user for transmission of a segment. Thus, \acp{BN} are associated with 
active users, \acp{SN} with slices in the frame, and edges with encoded segments.
A \ac{BN} corresponding to a type-$h$ user is called a type-$h$ \ac{BN}.
An edge incident on a type-$h$ \ac{BN} (and then corresponding to a type-$h$ segment) is called a type-$h$ edge. The number of edges
connected to
a BN or SN is the node degree. Therefore, a burst
encoded via the code $\mathscr{C}_h$ is represented as
a degree-$n_h$ \ac{BN}, and a slice where $d$
segments collide as a degree-$d$ \ac{SN}.

On the receiver side, according to the channel model introduced in Section~\ref{subsection:channel_model} segments experiencing collisions
do not provide any information while segments received in clean slices are received reliably. 
Hence, after all active users have transmitted their encoded segments in the \ac{MAC} frame, any BN may be thought as
connected to ``known'' edges and to ``unknown'' ones so that some of its encoded
segments are known, and the others unknown. At the generic BN (say of type $h$), erasure
decoding of code $\mathscr{C}_h$ may allow to recover some of the unknown encoded
segments. This enables to subtract the interference contribution of the newly
recovered encoded segments from the symbol in the corresponding slice. If $d-1$ segments
that collided in a SN of degree $d$ have been recovered by its neighboring BNs, the
remaining segment becomes known. The interference subtraction process combined with local decoding at the BNs
proceeds iteratively, i.e., cleaned slices may allow solving other collisions. 

Note that
this procedure is equivalent to iterative decoding of doubly-generalized low-density
parity-check (D-GLDPC) codes over the erasure channel \cite{paolini10:random}, where the variable
nodes are generic linear block codes and the check nodes are single parity-check (SPC) codes. 
Then, under the assumptions stated in Section~\ref{subsection:channel_model} the
\ac{IIS} process admits a representation as a message-passing procedure
along the edges of the above-introduced graph. Note also that the bipartite graph is not a priori known to the decoder, 
which ``discovers'' it during the iterative decoding process, based on the information available in each cleaned slice as discussed in Section~\ref{subsection:encoding_decoding}.

\begin{figure}[!t]
\begin{center}
\subfigure[Iteration $1$]{\includegraphics[width=0.4\columnwidth,draft=false]{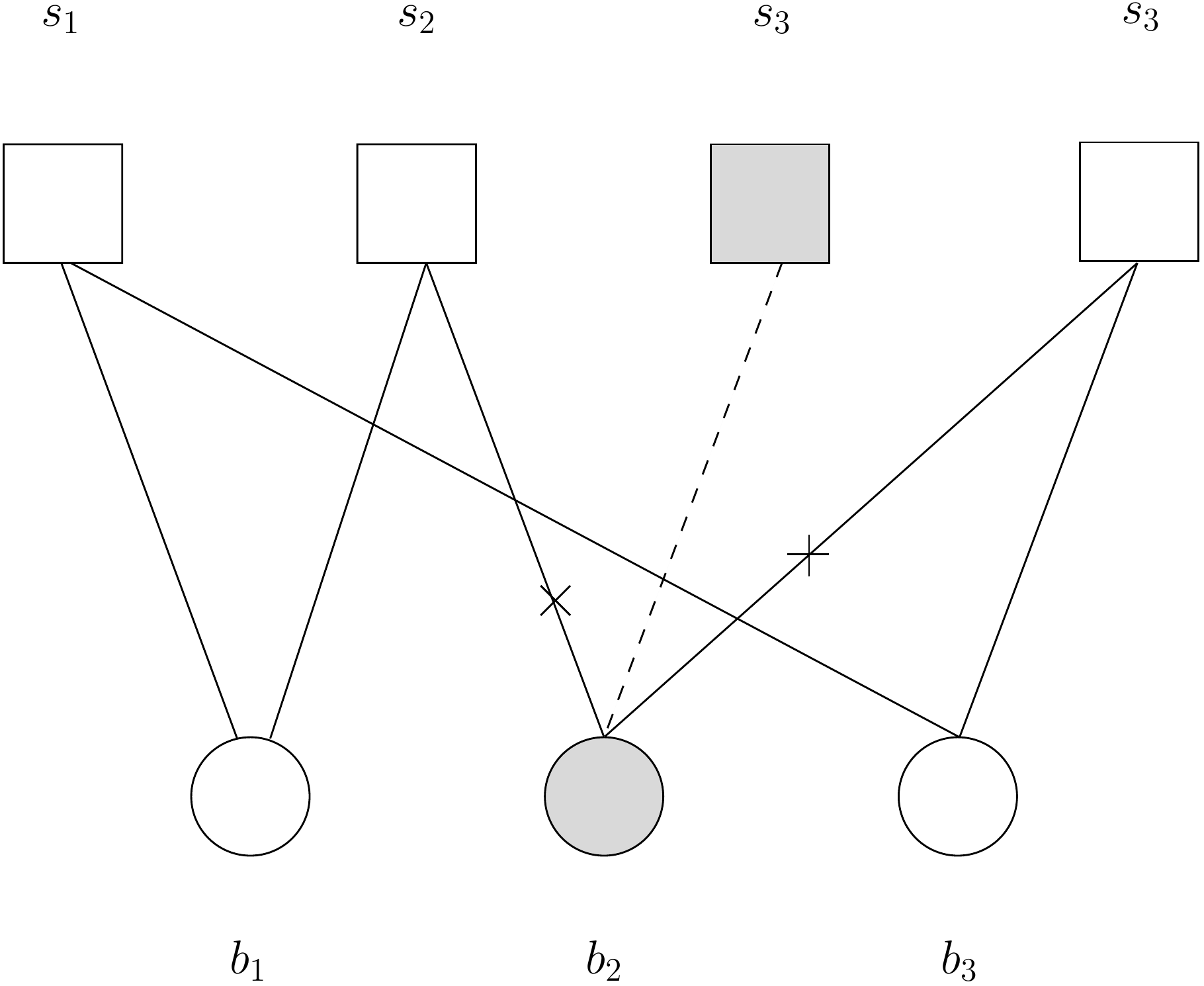}\label{fig:example_TIT_BP_i1}} \hfill
\subfigure[Iteration $2$]{\includegraphics[width=0.4\columnwidth,draft=false]{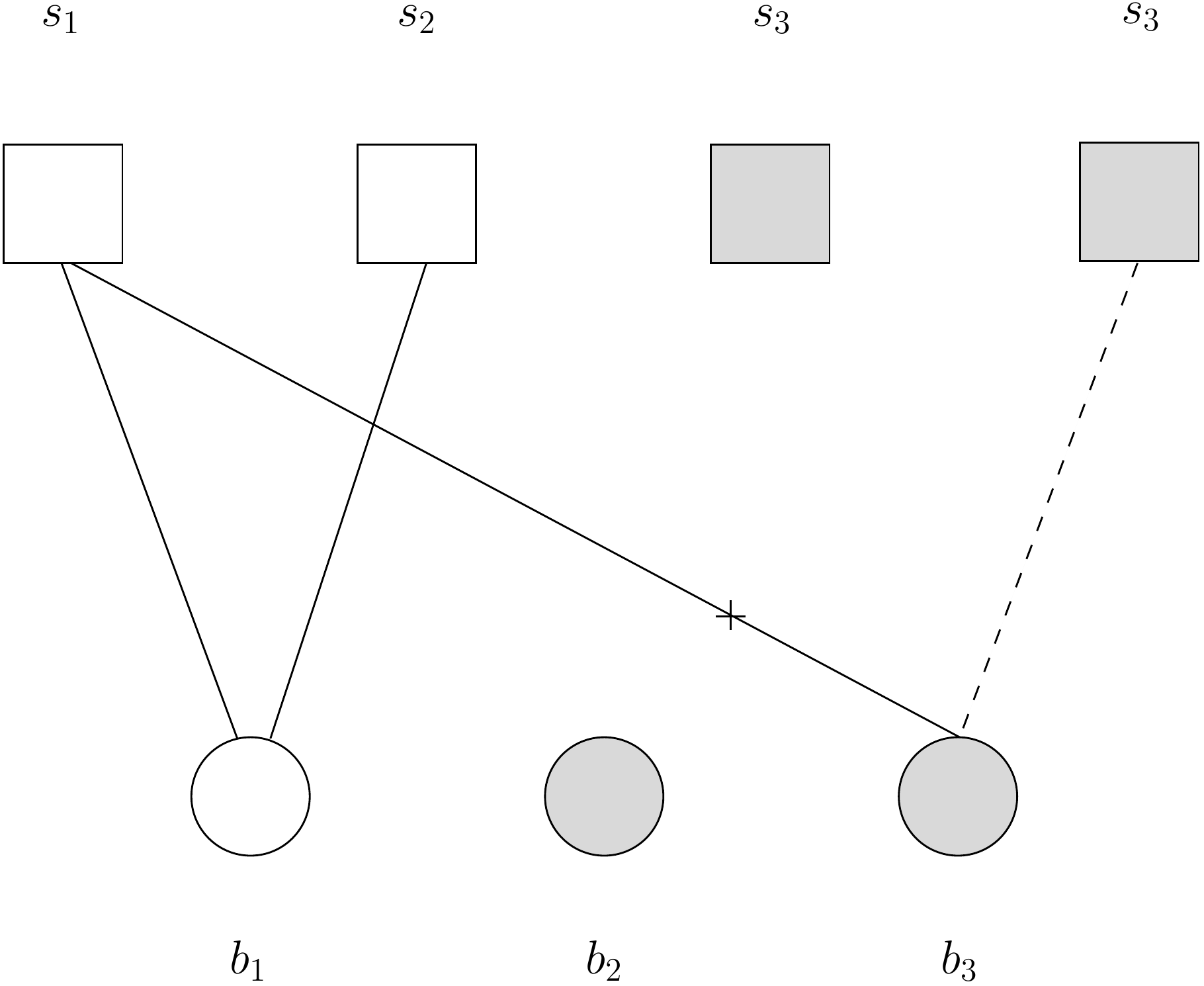}\label{fig:example_TIT_BP_i2}}
\subfigure[Iteration $3$]{\includegraphics[width=0.4\columnwidth,draft=false]{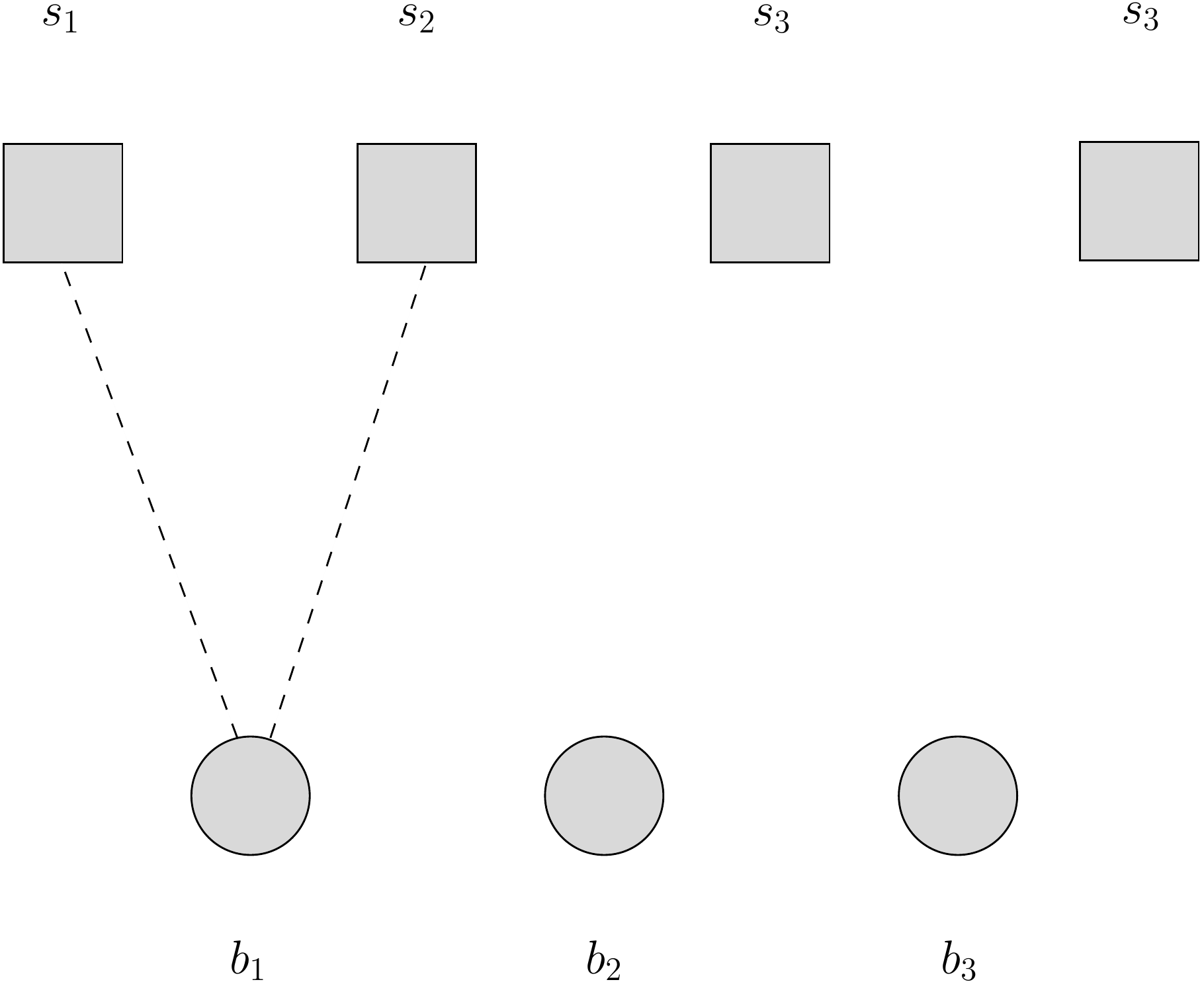}\label{fig:example_TIT_BP_i3}}
\end{center}
\caption{Example of graph representation of the \ac{IIS} process.}\label{fig:example_TIT_BP}
\end{figure}

\medskip
\begin{example}
In Fig.~\ref{fig:example_TIT}(a), an example of a \ac{MAC} frame composed by $4$ slices with $N_{\mathsf a}=3$ active users is displayed.
All three users adopt repetition codes ($k=1$), thus the number of slices corresponds to the number of slots in the frame.  The first and the third users encode their bursts with a $(2,1)$ code, whereas the second user employs a $(3,1)$ repetition code. Fig.~\ref{fig:example_TIT}(b) shows the corresponding bipartite graph model. According to the collision pattern of  Fig.~\ref{fig:example_TIT}(a), the first \ac{BN} is connected to the first and the second \acp{SN}, the second \ac{BN} connects to the last three \acp{SN}, and the third \ac{BN} is connected to the first and the fourth \acp{SN}. Fig.~\ref{fig:example_TIT_BP} illustrates how the graph model can be used to describe the iterative interference subtraction process. Observe  the collision pattern of  Fig.~\ref{fig:example_TIT}(a). The third slice contains an uncollided segment from the second user. Thus, the burst of the second user can be recovered and its interference contribution can be cancelled from the second and the fourth slice. This is shown through the graph model of Fig.~\ref{fig:example_TIT_BP_i1}, where a degree-$1$ \ac{SN} is detected ($s_3$), allowing the recovery of the second burst ($b_2$). Following the model of Fig.~\ref{fig:example_TIT_BP}, at the second iteration (Fig.~\ref{fig:example_TIT_BP_i2}) a degree-$1$ \ac{SN} is detected ($s_4$) allowing the recovery of the third burst ($b_3$). Finally, $b_1$ is recovered Fig.~\ref{fig:example_TIT_BP_i3} since it is connected to two \acp{SN} and they both have degree equal to $1$ in the residual graph. 
\end{example}


\subsection{Asymptotic Analysis of Iterative Interference Subtraction}\label{subsec:asymptotic_analysis_SIC}

In this subsection we analyze the evolution of the interference subtraction process in the \ac{CSA} scheme, for given $k$ and normalized population size $\alpha$, in the asymptotic case where $M$
(and correspondingly $N=\alpha M $) tends to infinity. We assume that 
\ac{MAP} erasure decoding is performed locally at each \ac{BN}.

We start by recalling the definition of information function of a linear block code
\cite{helleseth97:information}.
Consider an $(n,k)$ linear block code $\mathscr{C}$, where $n$ is the codeword length and $k$ the
code dimension, and let $\mathbf{G}$ be any generator matrix of $\mathscr{C}$. Then, the $g$-th
un-normalized information function of $\mathscr{C}$, denoted by $\tilde{e}_g$, is defined as the
summation of the ranks {of all} possible submatrices obtained selecting $g$ columns (with $0
\leq g \leq n$) out of $\mathbf{G}${, regardless their ordering}.

\medskip
\begin{lemma}\label{lemma:ql_pl-1}
Let $M \rightarrow \infty$ for constant normalized population size $\alpha$. Let $\tilde{e}^{(h)}_g$ be the $g$-th un-normalized information function for code $\mathscr C_h \in \mathcal C$.
At the $\ell$-th iteration of the {S}IC process, let $p_{\ell}$ be the probability that an edge is connected to a \ac{SN} associated with a segment where a collision persists. 
Moreover, let $q_{\ell}$ be the probability that an edge is connected to a \ac{BN} whose contribution of interference on the corresponding \ac{SN} cannot yet be cancelled, after \ac{MAP} decoding has been performed at each \ac{BN}. Then we have
\begin{align}\label{eq:q_i(p_i-1)}
q_{\ell} &= \frac{1}{\bar{n}}\sum_{h=1}^{\theta} \Lambda_h \sum_{t=0}^{n_h-1} p_{{\ell}-1}^t
(1-p_{{\ell}-1})^{n_h-1-t} \nonumber \\ 
& \times \left[ (n_h-t) \tilde{e}^{(h)}_{n_h-t}
- (t+1) \tilde{e}^{(h)}_{n_h-1-t} \right]  \, .
\end{align}
\end{lemma}
\begin{IEEEproof}
Exploiting the analogy between the \ac{IIS} process and iterative decoding over the erasure channel, $q_{\ell}$ is equal to the average extrinsic erasure probability (where the average is taken over the edges of the bipartite graph) outgoing from the \acp{BN} at the $\ell$-th \ac{IIS} iteration. It may be computed as the \ac{EXIT} function of the ``\ac{BN} decoder'' evaluated at the \emph{a priori} erasure probability incoming from the ``\ac{SN} decoder'', this latter probability being $p_{\ell-1}$. Denoting  the \ac{EXIT} function of the \ac{BN} decoder under \ac{MAP} decoding by $f_{\mathsf{b}}(\cdot)$, we~have
\begingroup
\allowdisplaybreaks
\begin{align}\label{eq:EXIT_BN_decoder}
q_{\ell} &= f_{\mathsf{b}}(p_{\ell-1}) \notag \\ 
&= \sum_{h=1}^{\theta} \lambda_h f_{\mathsf{b}}^{(h)}(p_{\ell-1})\, ,
\end{align}
\endgroup
where $\lambda_h$ is the probability that an edge is of type $h$ and where we have denoted by $f_{\mathsf{b}}^{(h)}(\cdot)$ the \ac{EXIT} function of a type-$h$ \ac{BN}.
It follows from \cite{Ashikhmin:AreaTheorem} that, if the linear block code $\mathscr C_h$ has no idle symbols, then the function $f_{\mathsf{b}}^{(h)}(\cdot)$ may be expressed as
\begin{align}\label{eq:EXIT_BN_h}
f_{\mathsf{b}}^{(h)}(p_{\ell-1}) &= \frac{1}{n_h}\sum_{t=0}^{n_h-1} p_{\ell-1}^t (1-p_{\ell-1})^{n_h-1-t} \nonumber \\
& \times [(n_h-t) \tilde{e}^{(h)}_{n_h-t} - (t+1) \tilde{e}^{(h)}_{n_h-1-t}] \, .
\end{align}
The theorem statement follows by incorporating \eqref{eq:EXIT_BN_h} into \eqref{eq:EXIT_BN_decoder} and by noting that  $\lambda_h=\frac{\Lambda_h n_h}{\bar n}$.
\end{IEEEproof}

\medskip
Equation \eqref{eq:q_i(p_i-1)} allows to update $q_{\ell}$ given $p_{\ell-1}$.
The dependence of $p_{\ell}$ on $q_{\ell}$ is instead stated by the following lemma. 

\medskip
\begin{lemma}\label{lemma:pl_ql}
Let $M \rightarrow \infty$ for constant normalized population size $\alpha$. Let $R$ be the rate of the scheme as defined in \eqref{eq:rate}.
At the $\ell$-th iteration of the {S}IC process, let $p_{\ell}$ and $q_{\ell}$ be defined as in the statement of Lemma~\ref{lemma:ql_pl-1}. Then we have
\begin{align}\label{eq:p_i(q_i)}
p_{\ell} = 1 - \exp\left\{ -\frac{\pi \alpha}{R}\, q_{\ell} \right\} \, .
\end{align}
\end{lemma}
\begin{IEEEproof}
For $0 \leq l \leq M$, the probability  $\Psi_l$ to receive $l$ encoded slices in a segment of the \ac{MAC} frame is given by
\[
\Psi_l={M\choose l}
\left(\frac{\bar{n}\pi \alpha}{kM}\right)^l\left(1-\frac{\bar{n}\pi \alpha}{kM}\right)^{M-l}\, .
\]
Defining $\Psi(x) =\sum_{l=0}^M \Psi_l x^l$ and letting $M\rightarrow \infty$ for constant $\alpha$, {yields}
\begin{align}\label{eq:Psix}
\Psi(x)=\exp \left\{ -\frac{\pi \alpha}{R}(1-x) \right\} \, .
\end{align}
Next, define the polynomial $\rho(x)=\sum_{l \geq 1} \rho_l x^{l-1}$, where $\rho_l$ is the probability that an edge in the bipartite graph is connected to a \ac{SN} of degree $l$.
Note that $\rho(x)$ is equivalent to the edge oriented degree distribution polynomial for the check nodes of an ordinary \ac{LDPC} code. We then have
\begin{align}\label{eq:rhox}
\rho(x) &= \frac{1}{\Psi'(1)} \, \frac{\mathrm{d} \Psi(x)}{\mathrm{d}x} \\
&= \Psi(x) \notag
\end{align}
and, from standard density evolution of \ac{LDPC} codes over the memoryless erasure channel,
$$
p_{\ell} = 1 - \rho(1-q_{\ell})
$$
which leads to \eqref{eq:p_i(q_i)}.
\end{IEEEproof}

\medskip
The right-hand side of \eqref{eq:p_i(q_i)} represents the \ac{EXIT} function of the \ac{SN} decoder. Hereafter, this function will be denoted by $f_\mathsf{s}(\cdot)$, so
\begin{align}\label{eq:fs_definition}
f_{\mathsf s}(q) = 1 - \exp\left\{ -\frac{\pi \alpha}{R}\, q \right\} \, .
\end{align}
From Lemma~\ref{lemma:ql_pl-1} and Lemma~\ref{lemma:pl_ql} we finally obtain a density evolution recursion for the \ac{IIS} process only involving the probability $p_{\ell}$.

\medskip
\begin{theorem}[Density evolution recursion for \ac{CSA}]
Let $M \rightarrow \infty$ for constant normalized population size $\alpha$. Let $R$ be the rate of the scheme as defined in \eqref{eq:rate} and $\tilde{e}^{(h)}_g$ be the $g$-th un-normalized information function for code $\mathscr C_h \in \mathcal C$. 
At the $\ell$-th iteration of the {S}IC process, let $p_{\ell}$ be defined as in the statement of Lemma~\ref{lemma:ql_pl-1}. Then we have
\begin{align}\label{eq:density_evolution}
p_{\ell} &= 1 - \exp \Bigg \{ -\frac{\pi \alpha}{k} \sum_{h=1}^{\theta} \Lambda_h
\sum_{t=0}^{n_h-1}  p_{\ell-1}^t (1-p_{\ell-1})^{n_h-1-t} \nonumber \\ 
& \times \left[ (n_h-t) \tilde{e}^{(h)}_{n_h-t} - (t+1) \tilde{e}^{(h)}_{n_h-1-t} \right] \Bigg\}
\end{align}
with starting point $p_0=1-\exp\{-\pi \alpha/R\}$. 
\end{theorem}
\begin{IEEEproof}
The recursion \eqref{eq:density_evolution} can be easily obtained as $p_{\ell} = ( f_{\mathsf s} \circ f_{\mathsf b} ) (p_{\ell-1})$, where 
$f_{\mathsf b}(\cdot)$ and $f_{\mathsf s}(\cdot)$ are defined in \eqref{eq:EXIT_BN_decoder} and \eqref{eq:fs_definition}, respectively, also noting that
from \eqref{eq:rate} we have $R \bar n = k$. The starting point of the recursion 
is equal to $p_0 = f_{\mathsf s}(1)$, i.e., to the average extrinsic erasure probability outgoing from the \ac{SN} decoder, when no \emph{a priori} information is 
available from the \ac{BN} decoder.
\end{IEEEproof}

\medskip
The density evolution recursion \eqref{eq:density_evolution} captures both the iterative cancellation of interference at the \acp{SN} and local \ac{MAP} decoding at the \acp{BN}. It may be specialized in the \ac{IRSA} case, in which all component codes in $\mathcal C$ are repetition codes.
This is expressed by the following corollary, in which we use the convention that the $h$-th component code is a length-$h$ repetition code (hence $n_h=h$) and that $\Lambda_1=0$.

\medskip
\begin{corollary}[Density evolution recursion for \ac{IRSA}]
Let $M \rightarrow \infty$ for constant normalized population size $\alpha$. Let $R$ be the rate of the scheme as defined in \eqref{eq:rate} and assume that code $\mathscr C_h \in \mathcal C$ is a length-$h$ repetition code, for $h \in \{2,\dots,\theta\}$. At the $\ell$-th iteration of the interference subtraction process, let $p_{\ell}$ be defined as in the statement of Lemma~\ref{lemma:ql_pl-1}. Then we have
\begin{align}\label{eq:density_evolution_irsa}
p_{\ell} = 1 - \exp \left\{ -\pi \alpha \sum_{h=2}^{\theta} h\, \Lambda_h\, p_{\ell-1}^{h-1} \right\}
\end{align}
with starting point $p_0=1-\exp\{-\pi \alpha/R\}$. 
\end{corollary}
\begin{IEEEproof}
The recursion \eqref{eq:density_evolution_irsa} follows directly from \eqref{eq:density_evolution} by observing that, when the code $\mathcal C_h$ is a length-$h$ repetition code, $k=1$ and the quantity $[(n_h-t) \tilde{e}^{(h)}_{n_h-t} - (t+1) \tilde{e}^{(h)}_{n_h-1-t}]$ is equal to zero for all $0 \leq t < h-1$ and is equal to $h$ for $t=h-1$.
\end{IEEEproof}

\medskip
For a given set $\mathcal C$ of component codes, a given \ac{p.m.f.} {\boldmath $\Lambda$} on $\mathcal C$, and a given normalized population size $\alpha$, the \emph{asymptotic threshold} of the \ac{CSA} access scheme,
denoted by $\pi^*=\pi^*(\mathcal{C},\mbox{\boldmath $\Lambda$},\alpha)$, is defined as 
$$
\pi^*(\mathcal{C},\mbox{\boldmath $\Lambda$},\alpha) := \sup \{ \pi \geq 0 \, | \, p_{\ell}\rightarrow 0 \textrm{ as } \ell\rightarrow\infty \}
$$
according to the recursion \eqref{eq:density_evolution}. The asymptotic threshold may also be defined in terms of the expected channel load, as $G^*(\mathcal{C},\mbox{\boldmath $\Lambda$}) := \alpha\, \pi^*(\mathcal{C},\mbox{\boldmath $\Lambda$},\alpha)$, this latter definition having the advantage to be independent of the normalized population size. 
In the asymptotic setting $M\rightarrow\infty$, for all $G<G^*(\mathcal{C},\mbox{\boldmath $\Lambda$})$
the throughput is $S=G$, i.e., all collisions are resolved even if packet retransmissions are forbidden. In this sense,
$G^*(\mathcal{C},\mbox{\boldmath $\Lambda$})$
represents the \emph{capacity} of the \ac{CSA} scheme on a slot-aligned collision channel without feedback conditional to
the specific choice of $\mathcal{C}=\{\mathscr{C}_1,\mathscr{C}_2,\dots,\mathscr{C}_{\theta}\}$ and
\mbox{\boldmath $\Lambda$}.

\begin{figure}[t]
\begin{center}
\includegraphics[width=0.9\columnwidth,draft=false]{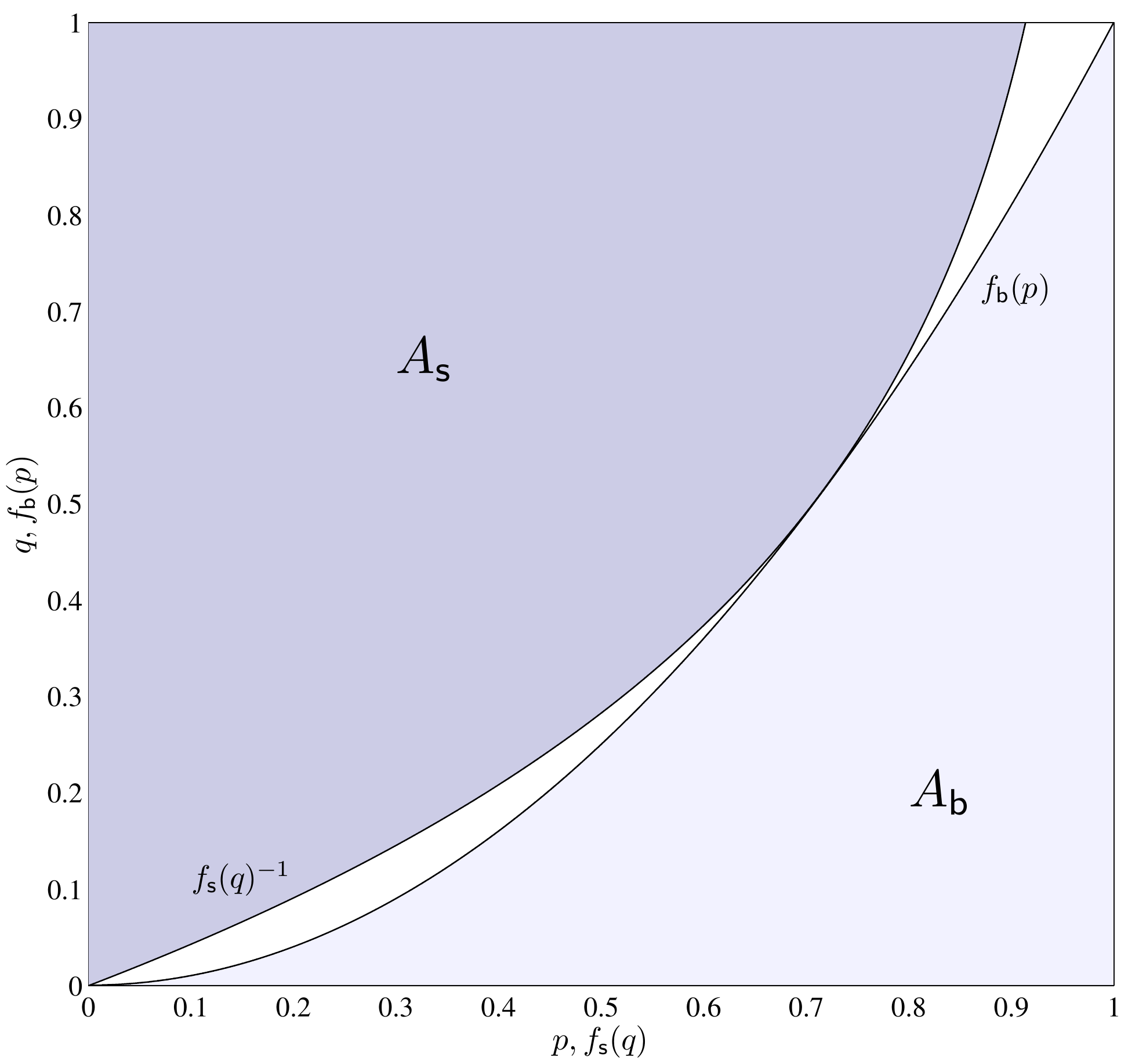}
\end{center}
\caption{\ac{EXIT} chart for a regular coded random access scheme employing a
rate-$1/3$ repetition code at each \ac{BN}, characterized by $G^*=0.816$.
}\label{fig:EXIT}
\end{figure}

The recursion defined by \eqref{eq:q_i(p_i-1)} and \eqref{eq:p_i(q_i)} can be
visualized in an \ac{EXIT} chart, which
displays  $f_{\mathsf{b}}(p)$ vs. $f_{\mathsf{s}}^{-1}(p)$.
An example of \ac{EXIT} chart for an \ac{IRSA} scheme in which $\mathcal C = \{ \mathscr C_1 \}$ where $\mathscr C_1$ is $(3,1)$
repetition code is provided in Fig.~\ref{fig:EXIT}. As the iteration index $\ell$
increases, the evolution of the pair of
probabilities $(p_{\ell},q_{\ell})$
traces a zig-zag pattern inside the tunnel between
the two curves. Whenever we operate the scheme below its capacity,
$G<G^*(\mathcal{C},\mbox{\boldmath $\Lambda$})$, the two curves do not intersect, leaving the tunnel
open.
This lets the pair of probabilities $(p_{\ell},q_{\ell})$ get arbitrarily close to the $(0,0)$ point. On the contrary, if
the scheme is operated above its capacity, $G>G^*(\mathcal{C},\mbox{\boldmath $\Lambda$})$, the two
curves
intersect (closing the tunnel) in a point $(\hat{p},\hat{q})$ with $\hat{p}>0, \hat{q}>0$,
and the \ac{IIS} process converges to a fixed point
corresponding to a non-zero residual erasure probability. 
As it was pointed out right after Example~\ref{example:encoding} in Section~\ref{subsection:encoding_decoding}, 
the performance of the \ac{CSA} scheme does not depend on the specific choice of the generator matrices for the $\theta$ component codes. Note, in fact, that the information functions $\tilde e_g^{(h)}$ in \eqref{eq:density_evolution}
are independent of the representation of code $\mathscr C_h$, $h\in \{1, \dots, \theta \}$, hence so are the \ac{EXIT} functions and the threshold $\pi^*(\mathcal{C},\mbox{\boldmath $\Lambda$},\alpha)$. The same holds for the performance of finite-length \ac{CSA} schemes, addressed in Section~\ref{sec:Results}.

\subsection{Stability of Iterative Interference Subtraction Collision-Free Point}

Autonomous difference equations such as \eqref{eq:density_evolution} are often 
analyzed as regard to the stability of their solutions or, as a particular case, of their fixed 
(steady-state equilibrium) points. In this subsection we study the stability of the fixed point $\hat p=0$ of \eqref{eq:density_evolution}, representing the
collision-free state. We remark that stability is here intended as \emph{convergence}, i.e., as the property of an equilibrium point $\hat x$ of 
a recursion $x_{\ell}=F(x_{\ell-1})$ to attract (in the sense of convergence as $\ell \rightarrow \infty$) 
the state $x_{\ell}$ when the initial state $x_0$ is perturbed from $\hat x$. Under this acceptation, a solution $x_{\ell} = \hat{x}$ $\forall \ell \geq 0$ of the difference equation $x_{\ell}=F(x_{\ell-1})$ is said to be
locally stable if there exists $\delta>0$ such that $|x_0 - \hat{x}| < \delta$ implies $|x_{\ell}-\hat{x}| \rightarrow 0$ as $\ell \rightarrow \infty$.
The following well-known result establishes a
necessary and sufficient condition for local stability.
\medskip
\begin{lemma}\label{lemma:stability}
A solution $x_{\ell} = \hat{x}$ $ \forall \ell \geq 0$ of a difference equation $x_{\ell}=F(x_{\ell-1})$,
where \mbox{$F:\mathbb{R} \mapsto \mathbb{R}$} is a differentiable function, is locally
stable if and only if $| F'(\hat{x}) | < 1$.
\end{lemma}

\medskip
The application of Lemma~\ref{lemma:stability} to \eqref{eq:density_evolution} yields the following~result.

\medskip
\begin{theorem}[Stability condition for \ac{CSA}]\label{proposition:stability}
For $h \in \{1,\dots,\theta\}$, let $\mathscr{C}_h$ be the $(n_h,k)$ linear block code with minimum distance $d_h \geq 2$ and without idle symbols,
employed with probability $\Lambda_h$ by the generic user to generate its encoded segments. Let $B^{(h)}_w$ be the number of
weight-$w$ codewords of $\mathscr{C}_h$. Moreover, let 
$$
r = \min_{ h \in \{ 1, \dots, \theta \} } \{d_h \} 
$$
and
$$
\mathcal{H}=\{h : d_h = r \} \, .
$$ 
If $r=2$, then the solution $p_{\ell} = \hat p = 0$ $\forall \ell \geq 0$ of recursion \eqref{eq:density_evolution} is locally stable if and only if
\begin{align}\label{eq:stability_p}
\pi < \frac{k}{2 \alpha \mathcal{B}_2}
\end{align}
where $\mathcal{B}_2=\sum_{h \in \mathcal H} \Lambda_h B^{(h)}_2$ is the expected number of weight-$2$ codewords in a code picked from $\mathcal C$.
Else, if $r \geq 3$, the fixed point $\hat p = 0$ of \eqref{eq:density_evolution} is
stable for any value of $G$.
\end{theorem}
\begin{IEEEproof}
Let us denote by $\mathbf G^{(h)}_g$ the generic $ k \times g $ matrix obtained by selecting $g$
columns in (any representation of) the generator matrix of code $\mathscr{C}_h$, irrespective of the
order of the $g$ columns, and by $\sum_{\mathbf G^{(h)}_g}$ the summation over all ${n \choose g}$ such
matrices. Moreover, let us define 
$$
a^{(h)}_t := (n_h-t) \tilde{e}^{(h)}_{n_h-t} - (t+1) \tilde{e}^{(h)}_{n_h-1-t}\, .
$$ 
We have:
\begingroup
\allowdisplaybreaks
\begin{align}\label{eq:f_prime_(0)_1}
& (f_\mathsf{s} \circ  f_\mathsf{b})'(0) \notag \\ 
& = \frac{\pi \alpha}{k} \sum_{h=1}^{\theta} \Lambda_h e^{-\frac{G\, a_0^{(h)}}{k}} \left[ a_1^{(h)} - (n_h-1) a_0^{(h)} \right] \notag \\
\, & \stackrel{\textrm{(a)}}{=}  \frac{\pi \alpha}{k} \sum_{h=1}^{\theta} \Lambda_h a_1^{(h)}  \notag \\
\, & = \frac{2\pi \alpha}{k}\sum_{h=1}^{\theta}
\Lambda_h\left[\frac{(n_h-1)\tilde{e}^{(h)}_{n_h-1}}{2}-\tilde{e}^{(h)}_{n_h-2}\right] \notag \\
\, & \stackrel{\textrm{(b)}}{=} \frac{2\pi \alpha}{k}\sum_{h=1}^{\theta} \Lambda_h\left[k {n_h \choose
n_h-2}-\tilde{e}^{(h)}_{n_h-2}\right] \notag \\
\, & = \frac{2\pi \alpha}{k}\sum_{h=1}^{\theta}
\Lambda_h\sum_{\mathbf{G}^{(h)}_{n_h-2}} \left( k-\mathrm{rank}(\mathbf{G}^{(h)}_{n_h-2}) \right) \notag \\
\, & \stackrel{\textrm{(c)}}{=} \left\{\begin{array}{lll}
\frac{2\pi \alpha}{k} \mathcal{B}_2 & \textrm{ if } & r=2 \\ 0 &
\textrm{ if } &r \geq 3 \end{array} \right.
\end{align}
\endgroup
In the previous equation list (a) and (b) follow from the hypothesis $r \geq 2$. In particular, (a) is due to $a_0^{(h)} = n_h \tilde{e}_{n_h} - \tilde{e}_{n_h-1} = n_h\, k - n_h\, k = 0 $ and
(b) to $\tilde e^{(h)}_{n_h-1} = k\, n_h$, both relying on $r \geq 2$. Moreover, (c) is due to $r \geq 2$ and to  \cite[Proposition 2]{paolini09:stability}. Inequality \eqref{eq:stability_p} now follows from
$| (f_\mathsf{s} \circ  f_\mathsf{b})'(0) |<1$.
\end{IEEEproof}

\medskip
The stability condition is a necessary, but in general not sufficient condition for successful
decoding in that, for given $\mathcal C$ and {\boldmath $\Lambda$}, values of the channel load may exist, fulfilling the bound \eqref{eq:stability_p} but which are above the \ac{CSA} capacity. This implies
\begin{align}\label{eq:stability_bound}
G^*(\mathcal{C},\mbox{\boldmath
$\Lambda$})\leq\frac{k}{2\mathcal{B}_2}
\end{align}
which will be referred to as the \emph{stability upper bound} and whose right-hand side will be denoted by $G^*_{\mathsf{sb}}(\mathcal{C},\mbox{\boldmath
$\Lambda$})$. Note that in
the \ac{IRSA} case ($k=1$) we have $\mathcal{B}_2=\Lambda_2$, where $\Lambda_2$ is the probability to select the
length-$2$ repetition code from the set $\mathcal C$, which yields\footnote{The stability condition for the \ac{IRSA} scheme appears in \cite[Eq. (7)]{Liva11:IRSA}.}
$$
G^*(\mathcal{C},\mbox{\boldmath $\Lambda$})\leq \frac{1}{2 \Lambda_2} \, .
$$ 

When $r=2$, \eqref{eq:stability_bound} may be achieved with equality, this situation being equivalent to the well-known \emph{flatness condition} for \ac{LDPC} codes \cite{shokrollahi00:capacity-achieving}.
This is the case, for example, when $\theta=1$ and the binary linear block code employed by
all users is a $(k+1,k)$ \ac{SPC} code, as stated by the following corollary.
\medskip
\begin{corollary}\label{prop:SPC}
Let $\mathcal C = \{ \mathscr C \}$ and the linear block code $\mathscr{C}$ employed by all users be a $(k+1,k)$ \ac{SPC}
code. Then
\begin{align}
G^*(\mathcal{C},\mbox{\boldmath
$\Lambda$}) = \frac{1}{k+1}\, .
\end{align}
\end{corollary}
\begin{IEEEproof}
If all users employ a $(k+1,k)$ \ac{SPC} code, then the stability bound \eqref{eq:stability_bound} becomes $G^*(\mathcal{C},\mbox{\boldmath $\Lambda$})\leq 1/(k+1)$. In order to prove that the bound is achieved with equality, it suffices to show that density evolution recursion \eqref{eq:density_evolution}, which assumes the simple form\footnote{This form follows from the duality property proved in \cite[Section~IV-E]{Ashikhmin:AreaTheorem}. }
\begin{align}\label{eq:density_evolution_SPC}
p_{\ell} = 1 - \exp \left\{- \frac{(k+1)\pi \alpha}{k} \left[ 1 - \left( 1- p_{\ell-1} \right)^k \right] \right\} \, ,
\end{align}
converges to $0$ as $\ell \rightarrow \infty$ for $\pi \alpha = 1 / (k+1)$. The result follows by observing that the function
$$
F(p) = 1 - \exp \left\{- \frac{1}{k} [ 1 - \left( 1- p \right)^k ] \right\} 
$$
fulfills $F'(0)=1$ (hence its graph is tangent to that of the function $I(p)=p$ in the $(0,0)$ point), $F'(p)=\exp \{- \frac{1}{k} [ 1 - ( 1- p )^{k} ] \}(1-p)^{k-1}>0$ for all $p\in[0,1)$, and
$F''(p) = - \exp \{- \frac{1}{k} [ 1 - ( 1- p )^{k} ] \} [(1-p)^{2k-2} + (k-1)(1-p)^{k-2}] < 0$ for all $p \in [0,1)$.
\end{IEEEproof}

\subsection{Asymptotic Analysis Under a Random Component Code Hypothesis}\label{sec:Appendix_random_hypothesis}

In the system model description provided in Section~\ref{sec:System_Model}, the generic user has been assumed to encode its $k$ information
segments via an $(n_h,k)$ binary linear block code, with minimum distance $d_h \geq 2$, picked randomly with
\ac{p.m.f.} {\boldmath $\Lambda$}$=\{ \Lambda_h \}_{h=1}^{\theta}$ from an ensemble of $\theta$ component codes. In this
subsection, we consider a slightly different setting. Specifically, we assume that the generic user randomly picks a codeword length $n_s>k$ from an ensemble
$\mathscr N = \{n_1,\dots,n_{s_{\max}}\}$ with \ac{p.m.f.} {\boldmath $\Lambda$}$ = \{\Lambda_{n_s} \}_{s=1}^{s_{\max}}$ and
encodes its $k$
segments through a binary $ k \times n_s $ generator matrix drawn randomly with uniform probability
from the set of all $ k \times n_s $ binary matrices with rank $k$ and representing $(n_s,k)$ linear block
codes without idle bits and with minimum distance at least $2$. We are interested in calculating
the expected asymptotic threshold for this scheme, where expectation is over all such generator matrices. The advantage of this \emph{random code hypothesis}
is that it allows to release the analysis from considering a specific set of $\theta$ codes.

With respect to the previous case, the definition \eqref{eq:rate} of the rate $R$ and the expressions \eqref{eq:Psix} and \eqref{eq:rhox} of $\Psi(x)$ and $\rho(x)$, respectively, remain unchanged provided the definition of $\bar{n}$ is
updated as $\bar{n}=\sum_{s=1}^{s_{\max}} \Lambda_{n_s} n_s$. Analogously, the recursion \eqref{eq:p_i(q_i)} is not
affected by
the random code hypothesis. On the other hand, the recursion \eqref{eq:q_i(p_i-1)} is updated as follows. Denote by $\mathsf{G}_{n_s,k}$ the ensemble of all
$ k \times n_s $ binary matrices with rank $k$ representing linear block codes without idle bits and
with minimum distance at least $2$, and by $\mathbb{E}_{\mathsf{G}_{n_s,k}}[\cdot]$ the expectation
operator over the set $\mathsf{G}_{n_s,k}$ (with a uniform probability measure). Then we have
\begin{align}\label{eq:q_i(p_i-1)_random_codes}
q_{\ell} &= \frac{1}{\bar{n}}\sum_{s=1}^{s_{\max}} \Lambda_{n_s} \sum_{t=0}^{n_s-1} p_{\ell-1}^t (1-p_{\ell-1})^{n_s-1-t} \notag \\
&\times \left[ (n_s-t) \mathbb{E}_{\mathsf{G}_{n_s,k}} \left[ \tilde{e}_{n_s-t} \right] - (t+1)
\mathbb{E}_{\mathsf{G}_{ n_s,k }} \left[ \tilde{e}_{n_s-1-t} \right] \right] 
\end{align}
where again $\bar{n}=\sum_{s=1}^{s_{\max}} \Lambda_{n_s} n_s$. For $0<k<n_s$ and $0 \leq g \leq n_s$, the expected $g$-th unnormalized information 
$\mathbb{E}_{\mathsf{G}_{n_s,k}} \left[ \tilde{e}_g \right]$ may be calculated using results
from \cite{paolini10:random}, in particular as
\begin{align}
\mathbb{E}_{\mathsf{G}_{n_s,k}} \left[ \tilde{e}_g \right] = {n_s \choose g} \sum_{u=1}^{\min\{k,g\}} u\,
\frac{K(k,n_s,g,u,k)}{J(k,n_s,k)}
\end{align}
where $J(k,n_s,k)$ denotes the number of $ k \times n_s$ binary matrices with rank $k$, without all-zero columns and
without independent columns,\footnote{In this context, a column is called ``independent'' when it is linearly independent of all the other matrix columns.} and where $K(k,n_s,g,u,k)$ is the number of $ k \times n_s $
binary matrices with rank $k$, without all-zero columns, without independent columns and such that their left-most $g$
columns have rank $u$. The functions $J(m,n,r)$ and $K(m,n,g,u,r)$ may be evaluated recursively, as detailed in 
\cite[Theorem~4]{paolini10:random} and \cite[Theorem~5]{paolini10:random}, respectively.

Density evolution recursion for \ac{CSA} under the random code hypothesis is then given by
\begin{align}\label{eq:density_evolution_random}
p_{\ell} &= 1 - \exp \Bigg \{  -\frac{\pi \alpha}{k} \sum_{s=1}^{n_s} \Lambda_{n_s}
\sum_{t=0}^{n_s-1} p_{\ell-1}^t  (1-p_{\ell-1})^{n_s-1-t} \notag \\
& \times \left[ (n_s-t) \mathbb{E}_{\mathsf{G}_{n_s,k}} \left[ \tilde{e}_{n_s-t} \right] - (t+1) \mathbb{E}_{\mathsf{G}_{ n_s,k }} \left[ \tilde{e}_{n_s-1-t} \right] \right]  \Bigg{ \} }
\end{align}
with starting point $p_0=1-\exp\{-\pi \alpha/R\}$. For given $\mathscr N = \{n_1,\dots,n_{s_{\max}}\}$ and {\boldmath $\Lambda$}$ = \{\Lambda_{n_s} \}_{s=1}^{s_{\max}}$, 
the expected asymptotic threshold of the \ac{CSA} scheme under the random code hypothesis,
denoted by $G^*=G^*(\mathscr{N},\mbox{\boldmath $\Lambda$})$, is defined as the supremum of the ensemble
of all $G\geq 0$ such that $p_{\ell}\rightarrow 0$ as $\ell\rightarrow\infty$ in recursion \eqref{eq:density_evolution_random}.

Using a proof technique analogous to that of Theorem~\ref{proposition:stability}, it is easy to
show that the stability upper bound is again given by \eqref{eq:stability_bound}, where now
$\mathcal{B}_2=\sum_{s=1}^{s_{\max}} \Lambda_{n_s} \mathbb E_{\mathsf G_{n_s,k}} \left[ B_2 \right]$ and
\begin{align*}
\mathbb E_{\mathsf G_{n_s,k}} \left[ B_2 \right]  = \!{n_s \choose 2}\!\! \left(k-\!\!\sum_{u=1}^{\min\{k,n_s-2\}}\!\!u\,
\frac{K(k,n_s,n_s-2,u,k)}{J(k,n_s,k)}\right)
\end{align*}
is the expected number of weight-$2$ codewords of an $(n_s,k)$ linear block code whose generator matrix is picked uniformly at
random in the set $\mathsf{G}_{n_s,k}$.

\section{Capacity Limits of \ac{CSA} Schemes}\label{sec:Capacity}

In this section, we develop an upper bound on the capacity of the \ac{CSA} scheme, for a given rate $R$.  The upper bound is established in the following theorem.

\medskip
\begin{theorem}\label{theorem:capacity_bound}
For $0<R \leq 1$, let $\mathbb{G}{(R)}$ be the unique positive solution of the
equation
\begin{align}\label{eq:barG_eqn}
G=1-e^{-G/R}
\end{align}
in $[0,1)$. Then, the capacity $G^*(\mathcal{C},\mbox{\boldmath
$\Lambda$})$ of the \ac{CSA} scheme fulfills
\begin{align}\label{eq:capacity_bound}
G^*(\mathcal{C},\mbox{\boldmath $\Lambda$}) \leq \mathbb{G}{(R)}
\end{align}
for \emph{any} choice of $\mathcal{C}=\{\mathscr{C}_1,\mathscr{C}_2,\dots,\mathscr{C}_{\theta}\}$ and
\mbox{\boldmath $\Lambda$} corresponding to a rate $R$.
\end{theorem}
\begin{IEEEproof} For given $\mathcal{C}=\{\mathscr{C}_1,\mathscr{C}_2,\dots,\mathscr{C}_{\theta}\}$ and \mbox{\boldmath $\Lambda$}, the evolution of the
probabilities $(p_{\ell},q_{\ell})$ is governed by the recursions
$q_{\ell}=f_{\mathsf{b}}(p_{{\ell}-1})$ and $p_{\ell}=f_{\mathsf{s}}(q_{{\ell}})$ in \eqref{eq:q_i(p_i-1)} and
\eqref{eq:p_i(q_i)}, for all $\ell \geq 1$ and with $q_1=f_{\mathsf{b}}(0)$. Let us denote the areas below the \ac{BN} and the \ac{SN} \ac{EXIT}
functions over the interval $[0,1]$ by
\begin{align*}
A_{\mathsf{b}}=\int_0^1 f_{\mathsf{b}}(p)\D p
\end{align*}
and
\begin{align*}
A_{\mathsf{s}}=\int_0^1 f_{\mathsf{s}}(q)\D q
\end{align*}
respectively. (These two areas are highlighted in the example \ac{EXIT} chart depicted in Fig.~\ref{fig:EXIT}.) A necessary and sufficient condition for successful decoding is represented by the existence of 
an ``open tunnel'' between the two curves in the \ac{EXIT} chart, which necessarily implies\footnote{Inequality \eqref{eq:AREA_CONDITION} is a necessary but not sufficient condition for successful decoding.}
\begin{equation}\label{eq:AREA_CONDITION}
A_{\mathsf{b}}+A_{\mathsf{s}} \leq 1\, .
\end{equation}
In particular, \eqref{eq:AREA_CONDITION} must be satisfied for $G=G^*(\mathcal{C},\mbox{\boldmath
$\Lambda$})$. The area below the \ac{SN} \ac{EXIT} function \eqref{eq:p_i(q_i)} is given by
\begin{equation}
A_{\mathsf{s}}=1+\frac{R}{G} e^{-\frac{G}{R}}-\frac{R}{G} \, . \label{eq:AREA_S}
\end{equation}
Moreover, the area below the \ac{BN} \ac{EXIT} function \eqref{eq:q_i(p_i-1)} is given~by
\begin{align}\label{eq:AREA_B}
A_{\mathsf{b}} &= \sum_{h=1}^{\theta} \lambda_h \int_0^1 f^{(h)}_{\mathsf{b}}(p) \D p \notag \\ 
& \stackrel{\mathrm{(a)}}{=} \sum_{h=1}^{\theta} \lambda_h \frac{k}{n_h} \notag \\
& \stackrel{\mathrm{(b)}}{=} R
\end{align}
where $\mathrm{(a)}$ follows from the Area Theorem
\cite{Ashikhmin:AreaTheorem} and holds under the assumption of \ac{MAP}
erasure decoding at the burst node\footnote{The Area Theorem states that the area below the
\ac{MAP} \ac{EXIT} function of a linear block code without idle symbols equals its
code rate.}, and where $\mathrm{(b)}$ is due to $\lambda_h = \frac{n_h\Lambda_h}{\bar{n}}$, to $\sum_{h=1}^{\theta} \Lambda_h =1$, and to \eqref{eq:rate}.
By incorporating \eqref{eq:AREA_S} and
\eqref{eq:AREA_B} in \eqref{eq:AREA_CONDITION} we {obtain}
$$
R + \frac{R}{G} \,e^{-G/R} \leq \frac{R}{G}
$$
which may be recast as
\begin{equation}\label{eq:R_bound}
R(G) \geq -\frac{G}{\log(1-G)} \, .
\end{equation}
Next, define $\mathbb{G}(R)$ as the unique solution in $(0,1]$ of \eqref{eq:barG_eqn}, yielding
$R(\mathbb{G})=-\mathbb{G}/\log(1-\mathbb{G})$. Since \eqref{eq:R_bound} must hold in particular for $G=G^*(\mathcal{C},\mbox{\boldmath $\Lambda$})$ and since the function $y=-x/\log(1-x)$, $x\in[0,1)$, is monotonically
decreasing, we obtain \eqref{eq:capacity_bound}.
\end{IEEEproof}

\medskip
Note that, while $G^*(\mathcal{C},\mbox{\boldmath $\Lambda$})$ depends on $R$ through $\mathcal{C}$
and \mbox{\boldmath $\Lambda$}, its upper bound $\mathbb{G}(R)$ depends solely on $R$. An alternative
proof of the upper bound \eqref{eq:capacity_bound} is proposed in Appendix~\ref{appendix:proof_2}.
It is manifest from the alternative proof that, for any rate $R$, the asymptotic throughput cannot exceed the value $\mathbb G(R)$ even if a ``genie-aided'' decoding approach, consisting of solving the linear system of equations via Gaussian elimination, is followed.


\section{Design and Analysis of \ac{CSA} Random Access Schemes}\label{sec:Results}

In this section, numerical results on \ac{CSA} access schemes are illustrated. The section is divided into two parts. 
The objective of the first part (Section~\ref{subsec:performance}) is to show that the asymptotic tools developed in Section~\ref{sec:CSA_Bipartite}
may confidently be used to design access schemes for a finite \ac{MAC} frame size. In the process, \ac{CSA} schemes based on 
simple codes of dimensions $k=2$ and $k=3$ are compared with \ac{IRSA} schemes. Purpose of Section~\ref{subsec:performance} is also to highlight the rate region in which \ac{CSA} schemes provide advantages over \ac{IRSA} ones and the rate region in which \ac{IRSA} protocols are preferable. The second part of the section (Section~\ref{subsec:approaching}) is devoted to the design 
of \ac{CSA} probability distributions approaching the bound established by Theorem~\ref{theorem:capacity_bound}.

\subsection{Performance Analysis of Finite-Length \ac{CSA} Schemes}\label{subsec:performance}

The analysis tool developed in Section~\ref{subsec:asymptotic_analysis_SIC} allows to calculate the threshold $G^*(\mathcal{C},\mbox{\boldmath $\Lambda$})$ for a given choice of the $\theta$ linear block component codes $\mathscr{C}_h$,
$h\in\{1,\dots,\theta\}$, and of the \ac{p.m.f.} {\boldmath $\Lambda$}. Analogously, the tool developed in Section~\ref{sec:Appendix_random_hypothesis} allows to evaluate the threshold $G^*(\mathscr{N},\mbox{\boldmath $\Lambda$})$ of a CSA scheme under the
random code hypothesis, for a given choice of the codeword lengths $n_s$,
$s\in\{1,\dots,s_{\max}\}$, and of the \ac{p.m.f.} {\boldmath $\Lambda$}. These tools can be exploited to derive optimal (in the sense of maximizing the threshold) probability distributions {\boldmath $\Lambda$} in the two cases.

\begin{table}[!]
\caption{\ac{IRSA} \acp{p.m.f.} {\boldmath $\Lambda$} with rates
$1/3$, $2/5$, and $1/2$, and \ac{CSA} \acp{p.m.f.} {\boldmath $\Lambda$} for $k=2,3$
with rates $1/3$, $2/5$, $1/2$, and $3/5$, under the random code hypothesis.}\label{table:random_code_ensembles}
\begin{center}
\begin{tabular}{rcccc}
\thickhline
\multicolumn{5}{c}{\textsf{IRSA}} \\
\hline
            & $R=1/3$       & $R=2/5$ & $R=1/2$       & \\
$(2,1)$ & $0.554016$ & $0.622412$ & $1.000000$  & \\  
$(3,1)$ & $0.261312$ & $0.255176$ &                      & \\
$(4,1)$ &                     & $0.122412$ &                      & \\
$(6,1)$ & $0.184672$ &                 &                      & \\
$G^*(\mathcal C, \mbox{\boldmath $\Lambda$})$ & $0.8792$ & $0.7825$ & $0.5000$ & \\ 
$G^*_{\mathsf{sb}}(\mathcal C, \mbox{\boldmath $\Lambda$})$ & $0.9025$ & $0.8033$ & $0.5000$ & \\
\thickhline
\multicolumn{5}{c}{\textsf{CSA} $k=2$, random component codes} \\
\hline
            & $R=1/3$         & $R=2/5$      & $R=1/2$       &  $R=3/5$ \\
$(3,2)$ & $0.259929$   & $0.304961$ &  & $0.666667$ \\ 
$(4,2)$ & $0.053247$   & $0.144152$ & $1.000000$  & $0.333333$ \\ 
$(5,2)$ & $0.447058$   &                     &  & \\ 
$(6,2)$ &                       & $0.347701$                     &  & \\ 
$(7,2)$ &                       & $0.203186$  &  & \\ 
$(11,2)$ & $0.105258$ &  &  & \\ 
$(12,2)$ & $0.134509$ &  &  & \\ 
$G^*(\mathscr{N},\mbox{\boldmath $\Lambda$})$ & $0.9034$ & $0.8185$ & $0.6556$ & $0.4091$ \\ 
$G^*_{\mathsf{sb}}(\mathscr{N},\mbox{\boldmath $\Lambda$})$ & $0.9035$ & $0.8185$ & $0.7500$ & $0.4091$ \\
\thickhline
\multicolumn{5}{c}{\textsf{CSA} $k=3$, random component codes} \\
\hline
            & $R=1/3$       & $R=2/5$      & $R=1/2$  &  $R=3/5$ \\
$(4,3)$ & $0.173572$ &                      & $0.045538$   &   \\ 
$(5,3)$ & $0.010699$ & $0.579066$  &   & $1.000000$   \\ 
$(6,3)$ & $0.183304$ &                      & $0.863386$   &   \\ 
$(7,3)$ & $0.361921$ &                      & $0.091076$   &   \\ 
$(8,3)$ & $0.025012$ &                      &                       &   \\ 
$(10,3)$ &   & $0.025606$   &   &   \\
$(11,3)$ &   & $0.395328$   &   &   \\  
$(18,3)$ & $0.245492$ &   &   &   \\ 
$G^*(\mathscr{N},\mbox{\boldmath $\Lambda$})$ & $0.9107$ & $0.8386$ & $0.6868$ & $0.5078$ \\ 
$G^*_{\mathsf{sb}}(\mathscr{N},\mbox{\boldmath $\Lambda$})$ & $0.9143$ & $0.8918$ & $0.9227$ & $0.5250$ \\
\thickhline
$\mathbb G(R)$ & $0.9405$ & $0.8926$ & $0.7968$ & $0.6758$ \\
\thickhline
\end{tabular}
\end{center}
\end{table}

\begin{table}[!]
\caption{\ac{CSA} \acp{p.m.f.} {\boldmath $\Lambda$} for $k=2$
with rates $1/3$, $2/5$, $1/2$, and $3/5$, for a specific choice of the component codes.}\label{table:specific_code_ensembles}
\begin{center}
\begin{tabular}{rcccc}
\thickhline
\multicolumn{5}{c}{\textsf{CSA} $k=2$, specific component codes} \\
\hline
            & $R=1/3$       & $R=2/5$      & $R=1/2$  &  $R=3/5$ \\
            $(3,2)\phantom{_{\mathrm{(a)}}}$ & $0.259929$   & $0.304961$ &  & $0.666667$ \\ 
$(4,2)_{\mathrm{(a)}}$ & $0.053247$   & $0.144152$ &  &  \\ 
$(4,2)_{\mathrm{(b)}}$ &    &  & $1.000000$  & $0.333333$ \\ 
$(5,2)_{\mathrm{(a)}}$ & $0.259293$   &                     &  & \\ 
$(5,2)_{\mathrm{(b)}}$ & $0.098353$   &                     &  & \\ 
$(5,2)_{\mathrm{(c)}}$ & $0.089412$   &                     &  & \\ 
$(6,2)\phantom{_{\mathrm{(a)}}}$ &                       & $0.347701$                     &  & \\ 
$(7,2)\phantom{_{\mathrm{(a)}}}$ &                       & $0.203186$  &  & \\ 
$(11,2)\phantom{_{\mathrm{(a)}}}$ & $0.105258$ &  &  & \\ 
$(12,2)\phantom{_{\mathrm{(a)}}}$ & $0.134509$ &  &  & \\ 
$G^*(\mathcal{C},\mbox{\boldmath $\Lambda$})$ & $0.9030$ & $0.8229$ & $0.6793$ & $0.4286$ \\ 
$G^*_{\mathsf{sb}}(\mathcal{C},\mbox{\boldmath $\Lambda$})$ & $0.9241$  & $0.8311$  & $1.0000$ & $0.4286$ \\
\thickhline
\end{tabular}
\end{center}
\end{table}

Some optimized probability distributions, obtained applying the random code hypothesis, are shown in Table~\ref{table:random_code_ensembles}. Among the several possible algorithms available to find the global maximum of a nonlinear function,
differential evolution \cite{storn05:differential-book} has been used (with the exception of the $R=1/2$ \ac{IRSA} scheme,
for which the only possibility is that all users employ a $(2,1)$ repetition code). In the upper part of the
table,   \acp{p.m.f.} {\boldmath $\Lambda$} are reported for \ac{IRSA} schemes
with rates $1/2$, $2/5$, and $1/3$, while in the lower part  
\acp{p.m.f.} {\boldmath $\Lambda$} are detailed for \ac{CSA} schemes with $k=2$ and $k=3$ and with the same rates, with the inclusion of $R=3/5$. All distributions have been
optimized under the constraint that the smallest local rate allowed for each user is $1/6$.
For each \ac{IRSA} distribution the threshold $G^*(\mathcal{C},\mbox{\boldmath $\Lambda$})$ and the corresponding stability bound 
are shown. On the other hand, for each \ac{CSA} distribution both the threshold $G^*(\mathscr{N},\mbox{\boldmath $\Lambda$})$ under the 
random code hypothesis and the corresponding stability bound, are reported. For all rates $R$, the value of the capacity bound $\mathbb G(R)$ is shown in the last row of the table.

\begin{figure}[]
\begin{center}
\subfigure[]{\includegraphics[width=0.9\columnwidth,draft=false]{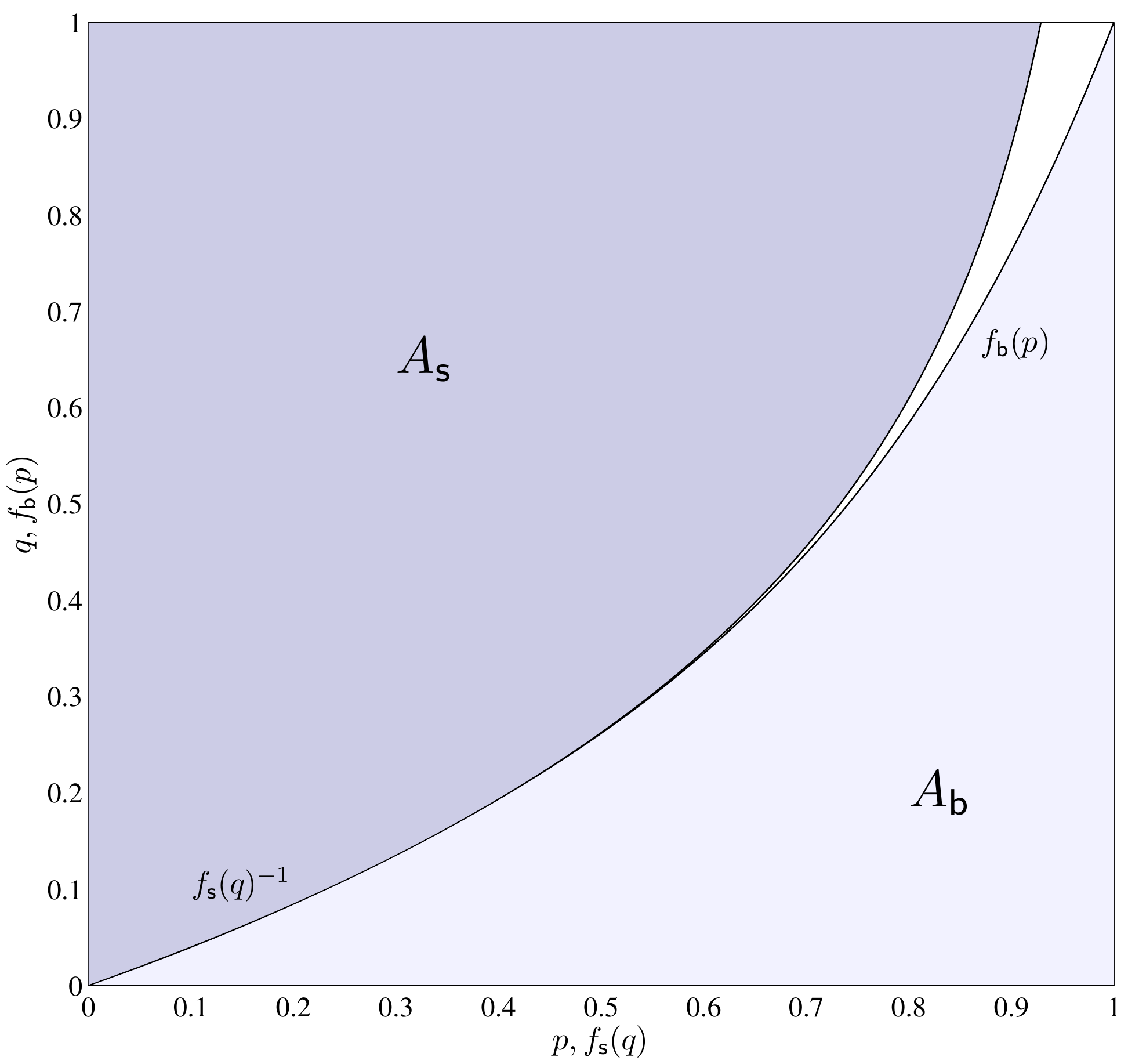}}
\subfigure[]{\includegraphics[width=0.9\columnwidth,draft=false]{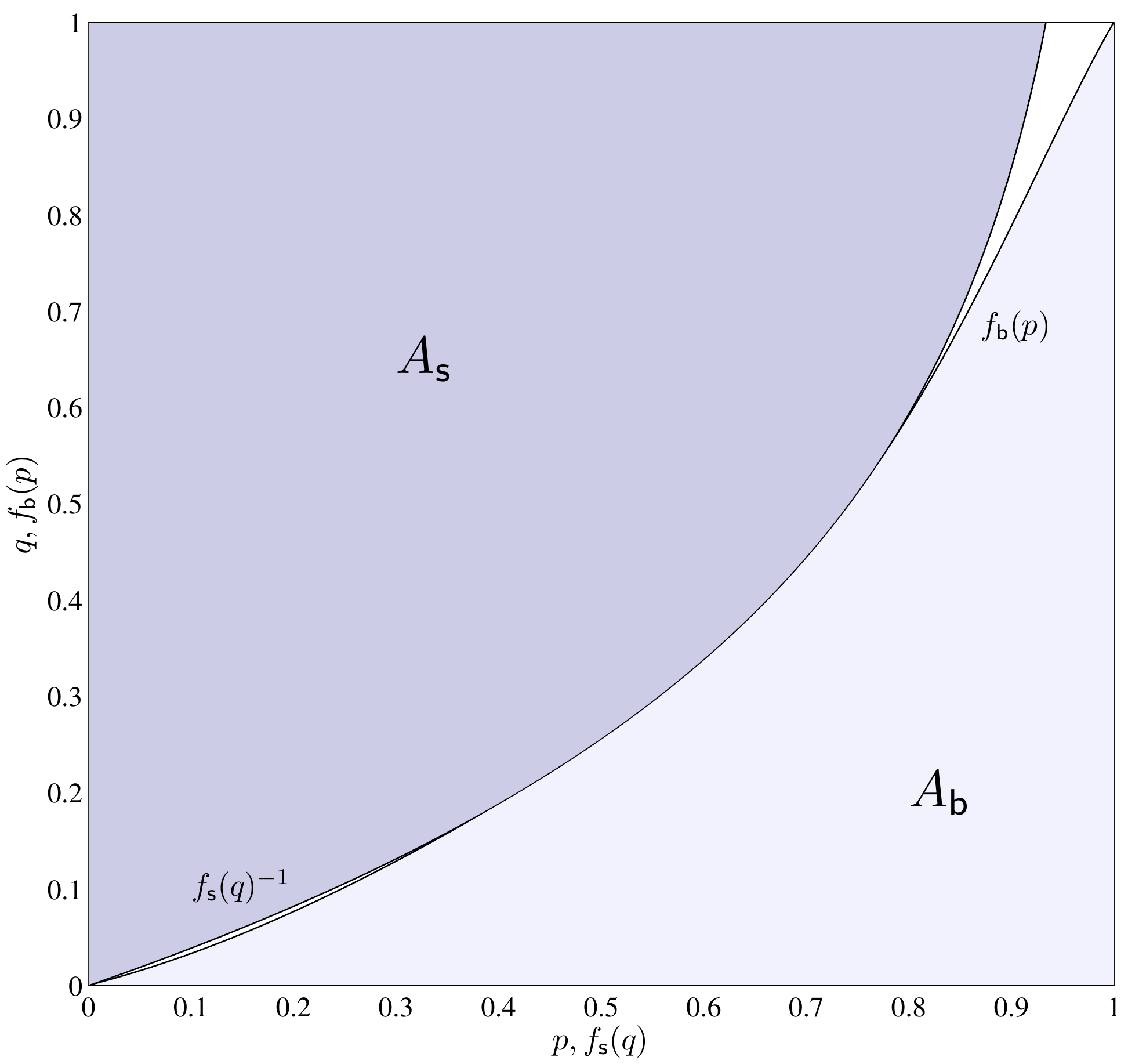}}
\end{center}
\caption{\ac{EXIT} charts for the $R=1/3$ \ac{IRSA} scheme in Table~\ref{table:random_code_ensembles} and for the $R=1/3$ \ac{CSA} scheme with $k=2$ in Table~\ref{table:specific_code_ensembles}. The \ac{EXIT} chart of the \ac{CSA} scheme is obtained with the choice \eqref{eq:component_codes_specific} of the component code generator matrices.}\label{fig:exit_areas}
\end{figure}

Table~\ref{table:specific_code_ensembles} shows the thresholds $G^*(\mathcal{C},\mbox{\boldmath $\Lambda$})$ for \ac{CSA} schemes with $k=2$ and characterized by the same \acp{p.m.f.} {\boldmath $\Lambda$} as the ones in Table~\ref{table:random_code_ensembles}, but for a specific choice of the component codes. More in detail, these thresholds have been obtained using linear block component codes generated by the following generator matrices:
\begingroup
\allowdisplaybreaks
\begin{align}\label{eq:component_codes_specific}
\mathbf{G}_{(3,2)} &= [110,011] \notag \\
\mathbf{G}_{(4,2)}^{\mathrm{(a)}} &= [1100,1111] \notag  \\
\mathbf{G}_{(4,2)}^{\mathrm{(b)}} &= [1100,0111] \notag  \\
\mathbf{G}_{(5,2)}^{\mathrm{(a)}} &= [11100,00111] \notag \\
\mathbf{G}_{(5,2)}^{\mathrm{(b)}} &= [11110,00011] \notag \\
\mathbf{G}_{(5,2)}^{\mathrm{(c)}} &= [11111,00011] \notag \\
\mathbf{G}_{(6,2)} &= [111000,001111] \notag \\
\mathbf{G}_{(7,2)} &= [1111000,0011111] \notag \\
\mathbf{G}_{(11,2)} &= [11110000000,00111111111] \notag \\
\mathbf{G}_{(12,2)} &= [111111110000,000001111111] \, . 
\end{align}
\endgroup
Note that this specific choice of the codes $\mathscr{C}_h$ leads to thresholds $G^*(\mathcal{C},\mbox{\boldmath $\Lambda$})$ which are either slightly larger than the corresponding ones in Table~\ref{table:random_code_ensembles} or practically coincident with them (as it is the case for the the rate-$1/3$ scheme). Also note that in \ac{CSA}, it is possible to combine different component codes having the same dimension and length. This is the case, for instance, of the $R=1/3$ scheme in Table~\ref{table:specific_code_ensembles} in which three different $(5,2)$ component codes are combined. The sum of the probabilities with which these three codes are picked by each user is equal to $0.447058$, the value in Table~\ref{table:random_code_ensembles} designed using the random code approach. For completeness, the \ac{EXIT} charts relevant to the $R=1/3$ \ac{IRSA} configuration in Table~\ref{table:random_code_ensembles} and to the $R=1/3$ \ac{CSA} scheme in Table~\ref{table:specific_code_ensembles} are depicted in Fig.~\ref{fig:exit_areas}(a) and Fig.~\ref{fig:exit_areas}(b), respectively.

As it was previously highlighted, \ac{CSA} allows to construct uncoordinated access schemes with any rate $0<R<1$, whereas only rates $0<R\leq1/2$ can be obtained with \ac{IRSA}, unless some users transmit their burst in the \ac{MAC} frame
with no repetition.\footnote{In case the set $\mathcal C$ for an \ac{IRSA} scheme includes repetition codes of length $1$, however, successful \ac{IIS} can never be
guaranteed due to the impossibility to subtract the interference of two bursts colliding in a slot and that have no replicas in other slots. As a consequence, density evolution
recursion \eqref{eq:density_evolution_irsa} will not converge to zero for any value of $\pi$ always
yielding \mbox{$G(\mathcal{C},\mbox{\boldmath $\Lambda$})^*=0$.}} (This is the reason for the optimized \ac{CSA} distributions of rate $R=3/5$ in Table~\ref{table:random_code_ensembles} have no \ac{IRSA} counterpart.) Furthermore, from Table~\ref{table:random_code_ensembles} and Table~\ref{table:specific_code_ensembles} we see that \ac{CSA} is capable to achieve better performance than \ac{IRSA}, in terms of asymptotic thresholds, over the whole range of rates $1/3 \leq R \leq 1/2$, and that the threshold values achieved by \ac{CSA} schemes are \emph{substantially} better than the ones achieved by \ac{IRSA} for values of $R$ that are close to $1/2$. For example, for $R=1/3$ a threshold $G^*(\mathscr{N},\mbox{\boldmath $\Lambda$})=0.9143$ is achieved by the best found \ac{CSA} scheme with $k=3$ (under the random code approach), whereas the best found \ac{IRSA} threshold is $G^*(\mathcal C, \mbox{\boldmath $\Lambda$})=0.8792$. For rate $R=1/2$ the improvement is much more pronounced, the threshold achieved by the best found \ac{CSA} scheme with $k=3$ (under the random code approach) being $G^*(\mathscr{N},\mbox{\boldmath $\Lambda$})=0.6868$ and the one achieved by \ac{IRSA} being $G^*(\mathcal C, \mbox{\boldmath $\Lambda$})=0.5000$.

For all tested values of $R$ we have observed improvements in terms of asymptotic threshold when the dimension $k$ of the component codes increases. This improvement becomes however almost negligible for low rates $R$ (equivalently, for high values of the excess energy $\Delta E = - 10 \, \log_{10} R$), a regime in which it is possible to design \ac{IRSA} schemes based on simple repetition codes, with thresholds very close to the upper bound $\mathbb G(R)$ (an example is represented by the distribution $\Lambda_1(x)$ that will be presented in Section~\ref{subsec:approaching}). Simplifying, we may conclude that for rates $R<1/3$ the \ac{IRSA} protocols should be preferred to \ac{CSA} ones, as their design is simpler and the gain provided by \ac{CSA} is limited. On the other hand, \ac{CSA} protocols are more appealing and effective than \ac{IRSA} ones in the range $1/3 \leq R \leq 1/2$ (where a higher energy efficiency is required) due to their better thresholds exhibited by the corresponding optimized distributions. \ac{CSA} schemes are a mandatory choice for $R>1/2$. For \ac{CSA} protocols, higher values of $k$ are effective in improving the threshold, as discussed further in Section~\ref{subsec:approaching}.

\begin{figure}[]
\begin{center}
\subfigure[]{\includegraphics[width=0.9\columnwidth,draft=false]{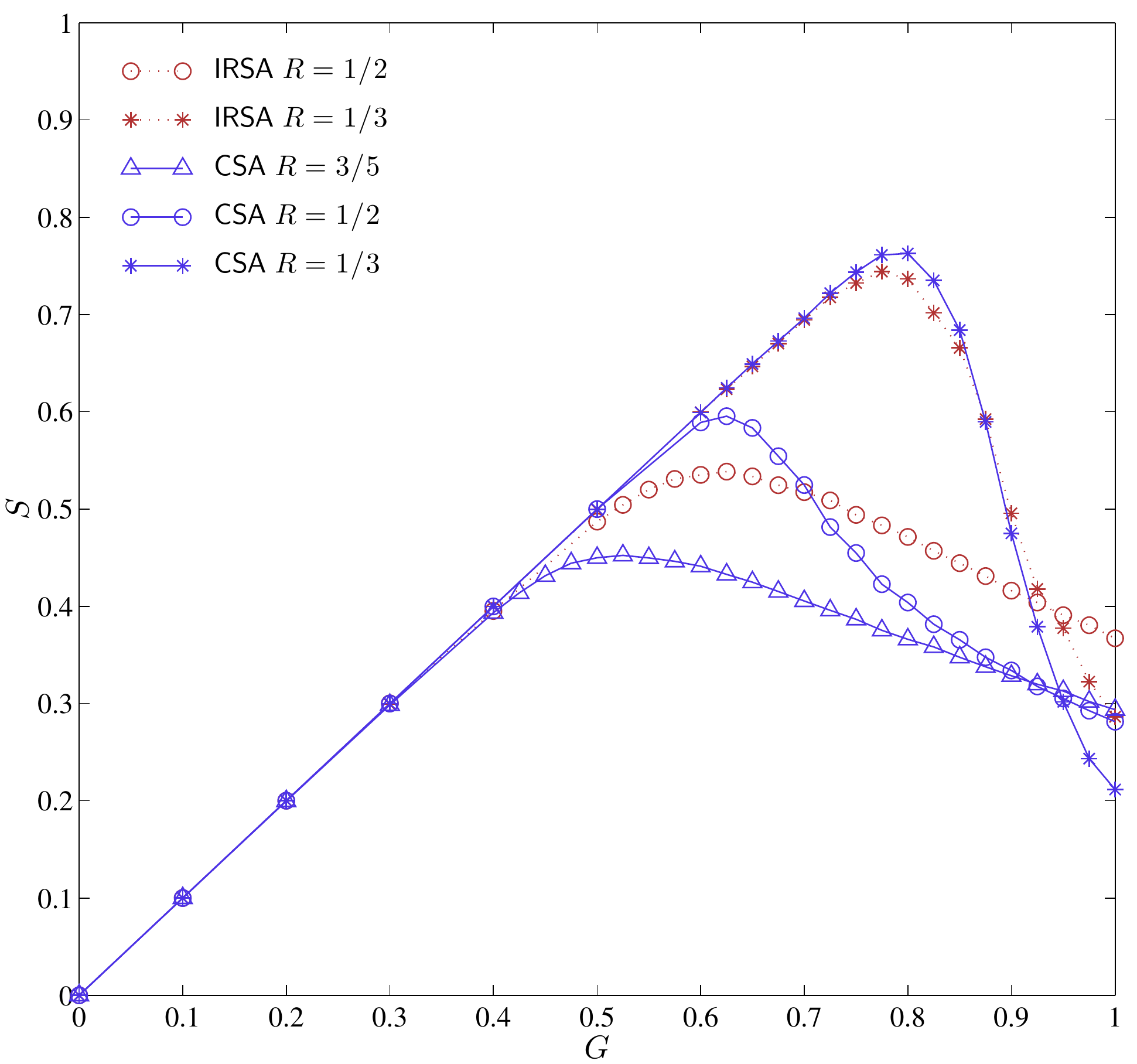}}
\subfigure[]{\includegraphics[width=0.9\columnwidth,draft=false]{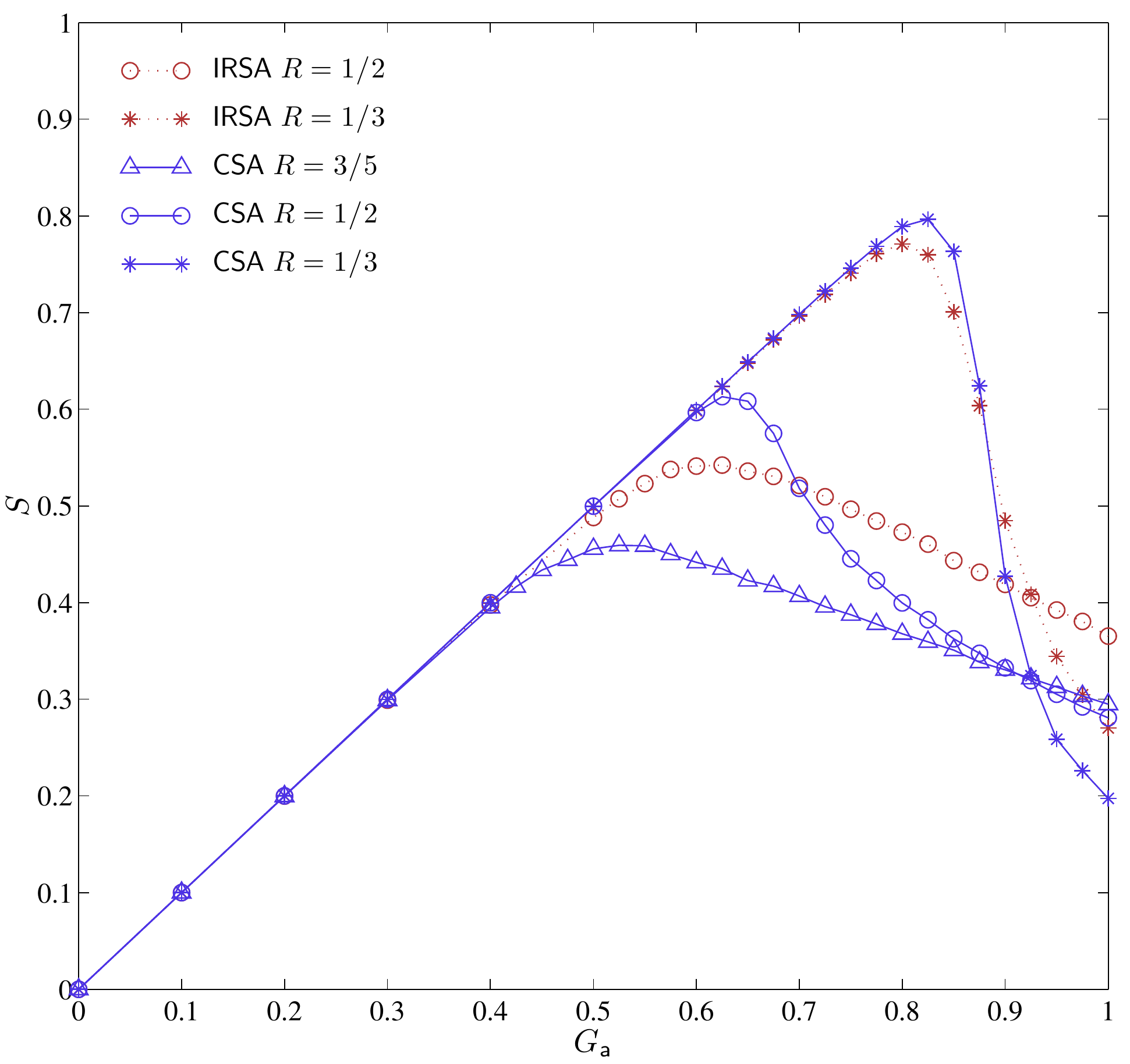}}
\end{center}
\caption{Throughput $S$ versus (a) the expected and (b) the instantaneous channel load for \ac{IRSA}
and \ac{CSA} ($k=2$) configurations with \acp{p.m.f.} $\mbox{\boldmath $\Lambda$}$ in Table~\ref{table:random_code_ensembles} and Table~\ref{table:specific_code_ensembles}, respectively. The linear block codes 
whose generator matrices are detailed in \eqref{eq:component_codes_specific} are employed for the \ac{CSA} case.}\label{fig:throughput}
\end{figure}

To validate our design approach based on the asymptotic analysis, we performed numerical simulations for 
finite frame size $M$ and user population size $N$. In Fig.~\ref{fig:throughput}(a) and Fig.~\ref{fig:throughput}(b), the throughput
curves without retransmissions of \ac{IRSA} schemes in Table~\ref{table:random_code_ensembles} and of \ac{CSA} ($k=2$) schemes in Table~\ref{table:specific_code_ensembles} are depicted as functions of the expected channel load $G$ and of the instantaneous channel load $G_{\mathsf a}$, respectively, for $R \in \{1/3,1/2,3/5\}$. In our simulations for the \ac{CSA} schemes, we used the linear block component codes generated by the generator matrices detailed in \eqref{eq:component_codes_specific}.
For the sake of fairness, we compared \ac{CSA} ($k=2$) and
\ac{IRSA} schemes for the same frame duration $T_{\mathsf{frame}}$ which implies $T_{\mathsf{slot}} = 2\, T_{\mathsf{segment}}$, i.e., a number of slices
twice the number of slots. Specifically, the simulations are for $T_{\mathsf{frame}}/T_{\mathsf{segment}}=1000$ slices and 
$T_{\mathsf{frame}}/T_{\mathsf{slot}}=500$ slots.\footnote{It should be considered that each segment has to be
encoded via a physical layer error correcting code before transmission on the \ac{MAC} channel, and
that the
physical layer code for \ac{CSA} is $k$ times shorter than the corresponding code for
\ac{IRSA}. Thus, \ac{CSA} may
require working at slightly higher \acp{SNR} than \ac{IRSA}, especially when short
segments (and then short physical layer codes) are used. This aspect is not captured by our collision channel model.} 
All simulations have been conducted for a population $N=20000$ users. The activation probability $\pi$ corresponding to each value of 
the expected channel load $G$ in Fig.~\ref{fig:throughput}(a) may be obtained as $\pi = G\, M / N$, while the number of active users for each value 
of the instantaneous load $G_{\mathsf a}$ in Fig.~\ref{fig:throughput}(b) as $N_{\mathsf a}=G_{\sf a} M$. We can observe how the trend of the peak throughput values measured in the finite
length case follow the same trend predicted by the asymptotic analysis. In particular, 
the slightly larger peak throughput exhibited by \ac{CSA} (for the specific choice of the component codes) for $R=1/3$ is in agreement 
with the thresholds reported in Table~\ref{table:random_code_ensembles} and Table~\ref{table:specific_code_ensembles}.

\begin{figure}[!t]
\begin{center}
\includegraphics[width=0.9\columnwidth,draft=false]{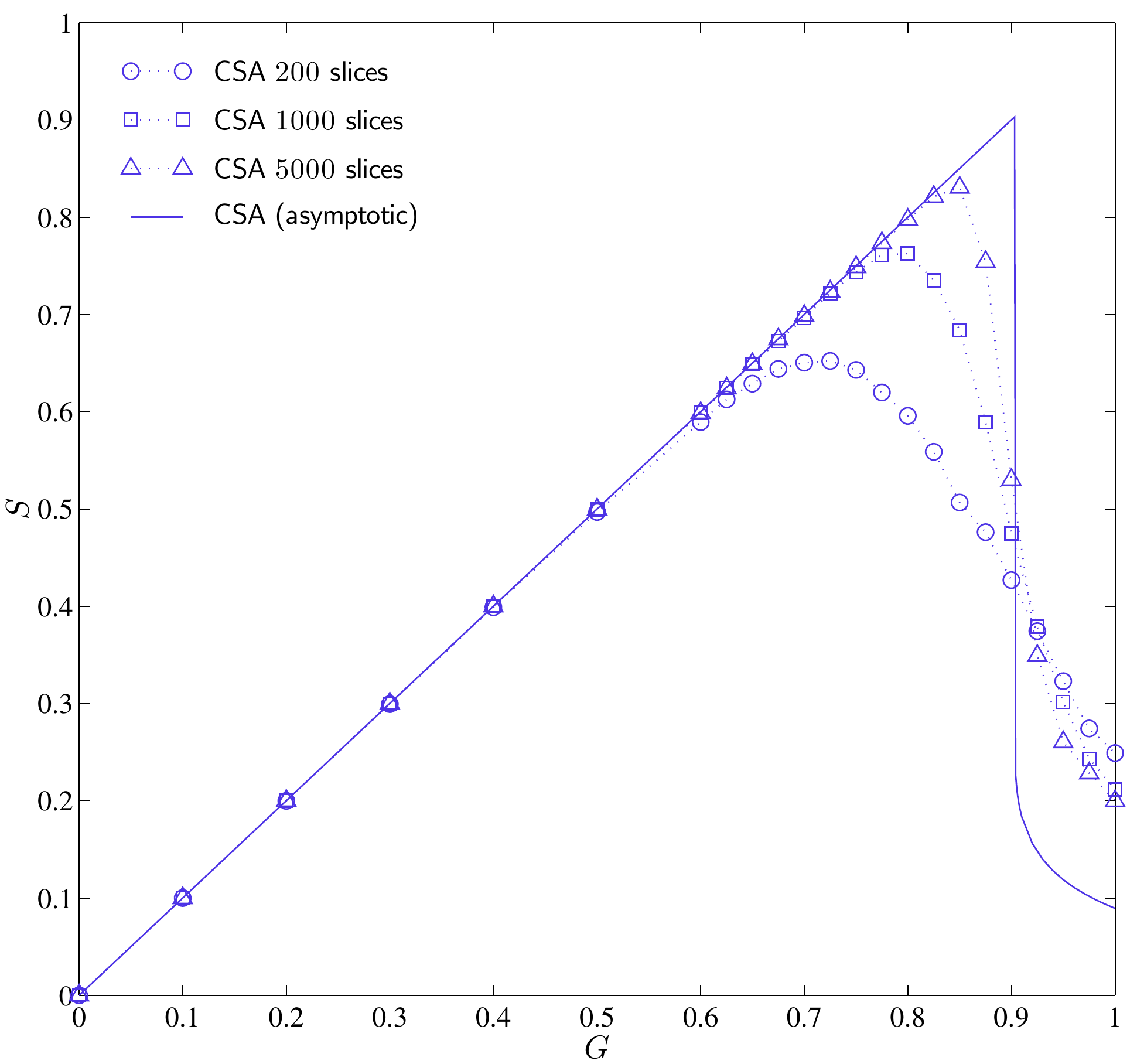}
\end{center}
\caption{Asymptotic throughput curve for the $R=1/3$ \ac{CSA} configuration in Table~\ref{table:specific_code_ensembles} and throughput curves for the same configuration under different lengths (in slices) of the \ac{MAC} frame. Population size $N=20000$.}\label{fig:throughput_var}
\end{figure}

For a given set $\mathcal C$ of component codes and a given \ac{p.m.f.} {\boldmath $\Lambda$}, the threshold $G^*(\mathcal C, \mbox{\boldmath $\Lambda$})$ represents
the asymptotic peak throughput of the corresponding \ac{CSA} scheme (in the limit where $M$ tends to infinity). In Fig.~\ref{fig:throughput_var} the asymptotic throughput curve
(versus the expected channel load $G$) of the rate $R=1/3$ \ac{CSA} scheme from Table~\ref{table:specific_code_ensembles}, with the component codes detailed in \eqref{eq:component_codes_specific}, is 
compared with the throughput curves obtained by numerical simulation for the same $(\mathcal C, \mbox{\boldmath $\Lambda$})$ pair, for $M=100$, $500$, and $2500$ slots 
(corresponding, for $k=2$, to $200$, $1000$, and $5000$ slices, respectively), always assuming $N=20000$ users. 
From this figure it is possible to appreciate how the curves for a finite number of slots tend to better and better fit the asymptotic curve as the number of slots increases.

\subsection{Approaching the Capacity Bound}\label{subsec:approaching}

In this subsection we consider the problem of designing \ac{CSA} configurations whose asymptotic 
thresholds approach the upper bound in Theorem~\ref{theorem:capacity_bound}. 
To do so, for a given $k$, a given
set $\mathcal{C}=\{\mathscr{C}_1,\mathscr{C}_2,\dots,\mathscr{C}_{\theta}\}$ of component codes, and a
given target rate $R$, we search (again via differential evolution optimization) the
distribution \mbox{\boldmath $\Lambda$} which maximizes $G^*(\mathcal{C},\mbox{\boldmath
$\Lambda$})$. In order  to limit the search space, we focus on
schemes based on codes of moderate-low length. We resort on a compact polynomial notation to specify
the developed \acp{p.m.f.} $\mbox{\boldmath $\Lambda$} = \{\Lambda_h \}_{h=1}^{\theta}$. This notation is introduced 
for each specific case before its usage.

\begin{figure}[t]
\begin{center}
\includegraphics[width=0.9\columnwidth,draft=false]{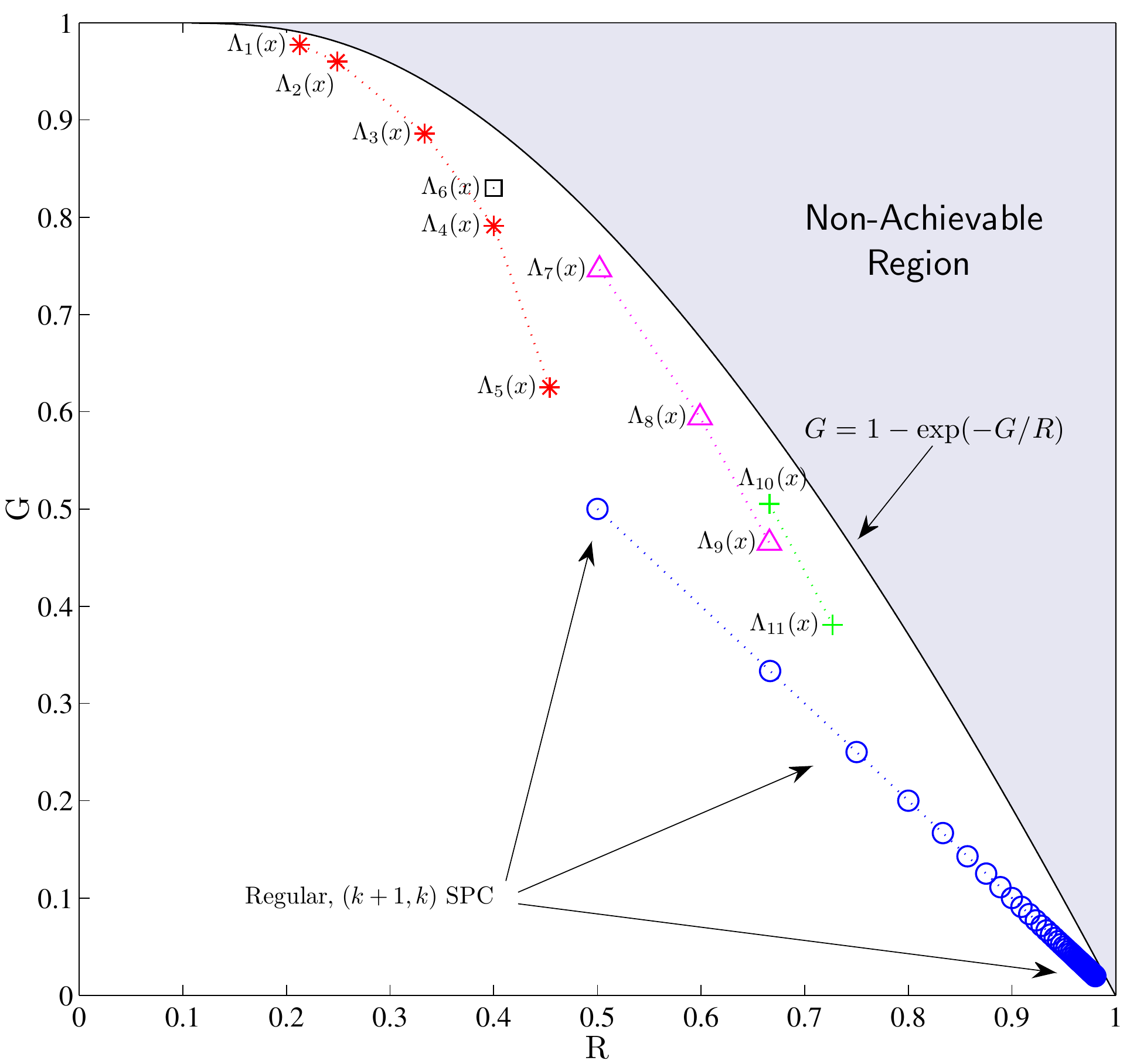}
\end{center}
\caption{Upper bound to the capacity vs. rate $R$. Thresholds $G^*$ are reported for
selected distributions. Distributions $\Lambda_i(x), i=1,\ldots, 5$  (\textcolor{red}{$\ast$}) are \ac{IRSA} configurations
based on repetition codes ($k=1$). Distribution $\Lambda_6(x)$ ({$\square$}) is based on
MDS codes with $k=2$. Distributions $\Lambda_i(x), i=7,8,9$
(\textcolor{magenta}{$\vartriangle$}) are based on \ac{MDS} codes with $k=3$. Distributions
$\Lambda_i(x), i=10,11$  (\textcolor{green}{$+$}) are based on \ac{MDS} codes with $k=4$.
Distributions based on $(k+1,k)$ \ac{SPC} codes are also displayed
(\textcolor{blue}{$\circ$}).}\label{fig:Capacity}
\end{figure}

Based on the observations in Section~\ref{subsec:performance}, we start by designing some low-rate \ac{IRSA} schemes, in which case we define $\Lambda(x)=\sum_h \Lambda_h x^h$, where $\mathscr{C}_h$ is the
$(h,1)$ repetition code. Selecting a rate $R=1/5$ and limiting the maximum length of the repetition component codes to $30$
(i.e., considering only repetition codes with rate down to $1/30$), we obtain the distribution
\begin{align*}
\Lambda_1(x) &= 0.494155x^2 +  0.159085x^3 +  0.107372x^4 \\ 
&+ 0.070336x^5 + 0.045493x^6 +  0.019898x^7 \\ 
&+ 0.024098x^{11} +  0.008636x^{12} +  0.005940x^{13} \\ 
&+  0.008749x^{15} + 0.002225x^{18} +  0.001261x^{20} \\
&+ 0.002607x^{22} +  0.008092x^{23} +  0.002287x^{24}\\
&+ 0.012274x^{25} +  0.002530x^{26} + 0.003094x^{27} \\ 
&+ 0.002558x^{28} +  0.005891x^{29} + 0.013419x^{30}
\end{align*}
whose threshold is $G^*(\mathcal C_1, \mbox{\boldmath $\Lambda$}_1)=0.977$.
The corresponding point on the $G$ versus $R$ plane is reported in
Fig.~\ref{fig:Capacity} and compared with the bound given by
Theorem~\ref{theorem:capacity_bound}. On the same chart the points corresponding to 
\ac{IRSA} distributions with different rates, denoted by $\Lambda_i(x)$ for $i\in\{2,\dots, 5\}$, are reported. Whereas for low
rates $R$ repetition-based configurations approach the bound quite tightly,
for rates close to $1/2$ they show visible losses. For example, the
distribution 
$$
\Lambda_5(x)=0.8x^2 +  0.2x^3
$$
(obtained constraining the maximum length of the component codes to $n_h=5$) is characterized by a rate $R=5/11$ and attains a
threshold $G^*(\mathcal C_5,\mbox{\boldmath $\Lambda$}_5)=0.625$, whereas $\mathbb{G}(5/11) = 0.843$. This effect if somehow expected since
in the limiting case of $R=1/2$, in which each user employs a $(2,1)$
repetition code, the corresponding threshold is limited to $0.5$.

\begin{figure}[t]
\begin{center}
\includegraphics[width=0.9\columnwidth,draft=false]{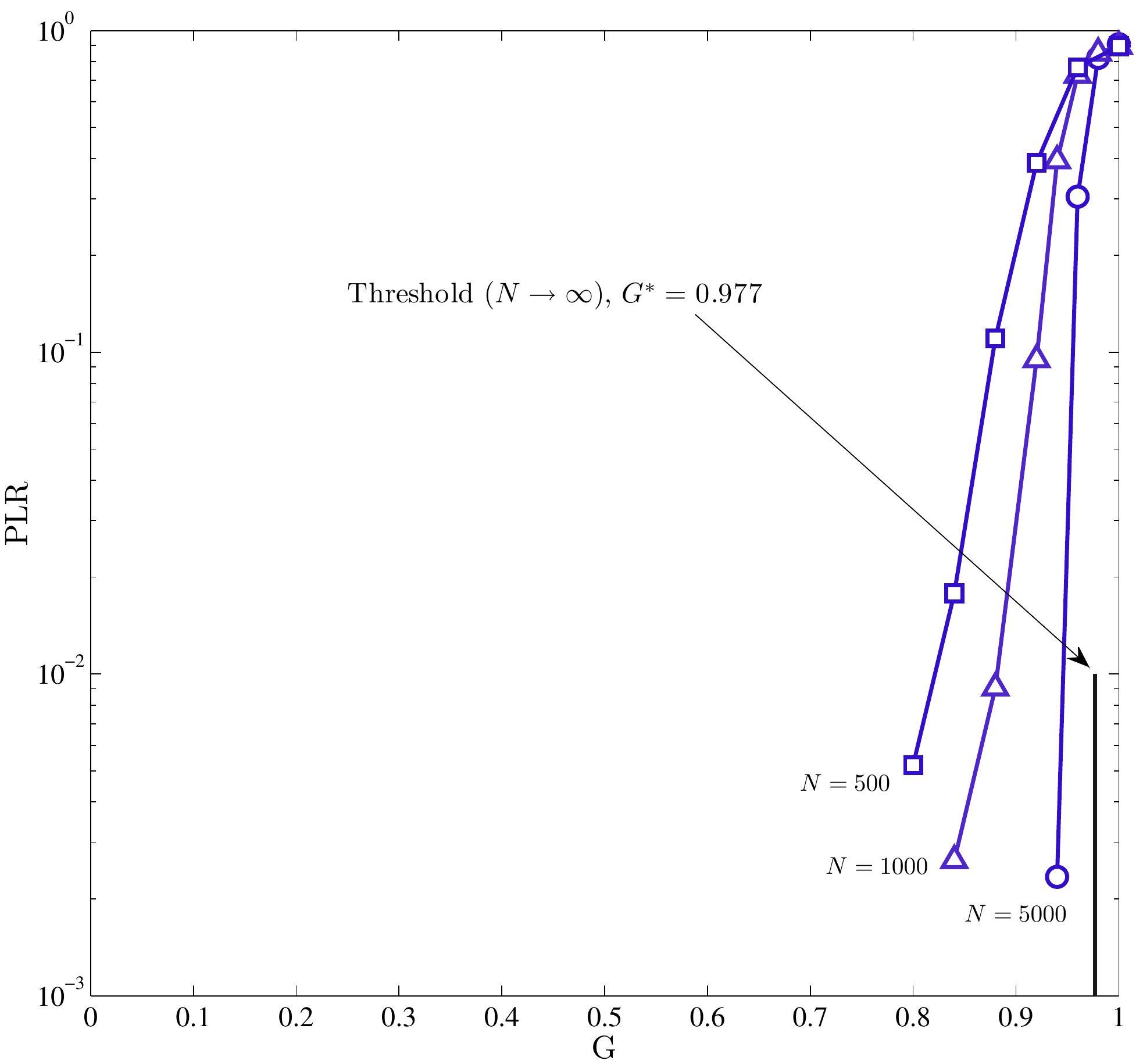}
\end{center}
\caption{Packet loss rate for the \ac{CSA} scheme based on the distribution $\Lambda_1(x)$. $N=5000, 1000,
500$, maximum iteration count set to $100$.}\label{fig:PLR}
\end{figure}

\begin{table*}[!t]
\caption{\ac{CSA} distributions for $k \in \{2,3,4 \}$ and corresponding asymptotic thresholds. All configurations are based on \ac{MDS} codes.}\label{table:MDS}
\begin{center}
\begin{tabular}{c|cccc}
\thickhline
        & $R$ & $\Lambda(x)$ & $G^*(\mathcal C, \mbox{\boldmath $\Lambda$})$ & $\mathbb G(R)$ \\
        \hline
$k=2$ & $0.400$      & $\Lambda_6(x)=0.276023x + 0.366641x^2 + 0.127979x^3 + 0.229357x^7$ & $0.830$ & $0.893$  \\ 
\hline
\multirow{3}{*}{$k=3$} & $0.502$ & $\Lambda_7(x)=0.3222 x + 0.2305 x^{2} + 0.0491 x^{4} + 0.3983 x^{5}$ & $0.746$ & $0.795$  \\
            & $0.599$ & $\Lambda_8(x)=0.2589 x + 0.4826 x^{2} + 0.2586 x^{3}$ & $0.594$ & $0.677$  \\
            & $0.667$ & $\Lambda_9(x)=0.5005 x + 0.4995 x^{2}$ & $0.465$ & $0.582$ \\
\hline
\multirow{2}{*}{$k=4$} & $0.667$ & $\Lambda_{10}(x)=0.1892 x + 0.6240 x^{2} + 0.1868 x^{3}$ & $0.505$ & $0.582$ \\
            & $0.727$ & $\Lambda_{11}(x)=0.5000 x + 0.5000 x^{2}$ & $0.381$ & $0.491$  \\
\thickhline
\end{tabular}
\end{center}
\end{table*}

Fig.~\ref{fig:PLR} shows the \ac{PLR} achieved by the scheme
employing the distribution $\Lambda_1(x)$ without retransmissions. The results have been
derived via Monte Carlo simulations for \ac{MAC} frames of size
$M=5000$, $1000$, and $500$ slots, and are compared with the
capacity of the scheme, $G^*(\mathcal C_1, \mbox{\boldmath $\Lambda$}_1)=0.977$. For the $M=5000$ case, a \ac{PLR}
close to $2\times 10^{-3}$ is achieved at a channel traffic $G=0.94\,\mathrm{[packets/slot]}$, only
$0.05\,\mathrm{[packets/slot]}$ away from the bound
established by Theorem~\ref{theorem:capacity_bound} ($\simeq0.99\,\mathrm{[packets/slot]}$).

As observed in Section~\ref{subsec:performance}, when the rate $R$ is not too low, e.g., $R \geq 1/3$, it becomes
convenient to adopt component codes with $k>1$. To this
purpose, we designed \ac{CSA} schemes where each $\mathscr{C}_h\in\mathcal{C}$ is an
\ac{MDS} code constructed on an appropriate non-binary finite field,\footnote{Imposing limits on $n_h$, this approach is realistic. For instance,
(generalized) Reed-Solomon codes on finite fields of moderate order may be
used.} for component code dimensions $k=2$, $3$, and $4$. It is assumed that each burst node locally adopts a
bounded-distance decoding strategy at each iteration,
consisting of recovering the lost encoded segments connected to it only if the current number of
its collision-free received segments is at least $k$. Under this
assumption, the \ac{EXIT} function of a \ac{BN} using an $(n_h,k)$
\ac{MDS} codes is given~by
\begin{align}
f_{\mathsf{b}}^{(h)}(p) =\sum_{l=0}^{k-1} {n_h-1 \choose l} (1-p)^{l}p^{n_h-l-1}.
\end{align}
In this case, we specify the \ac{CSA} \acp{p.m.f.} $\mbox{\boldmath $\Lambda$}$ via the compact polynomial notation $\Lambda(x)=\sum_h
\Lambda_h x^h$, where $\mathscr{C}_h$ is an $(h+k,k)$ \ac{MDS} code. The obtained \ac{CSA} distributions are reported in Table~\ref{table:MDS}.

For $k=2$ and $R=0.4$ we designed the distribution $\Lambda_6(x)$
characterized by a threshold $G^*(\mathcal C_6, \mbox{\boldmath $\Lambda$}_6)=0.830$. For the same rate, the best found \ac{IRSA} configuration for a maximum $n_h$ set to $10$ ($\Lambda_4(x)$ in Fig.~\ref{fig:Capacity}), 
achieves $G^*(\mathcal C_4, \mbox{\boldmath $\Lambda$}_4)=0.791$ while the best found \ac{CSA} ($k=2$) scheme based on binary codes in Table~\ref{table:specific_code_ensembles} achieves $G^*(\mathcal C, \mbox{\boldmath $\Lambda$})=0.8229$. Moving to the
moderate-rate regime, we observed that, also employing \ac{MDS} codes under bounded distance decoding as component codes, 
for the same rate $R$ the bound $\mathbb G(R)$ can be better approached resorting on \ac{CSA} distributions 
based on higher code dimensions $k$ (see Fig.~\ref{fig:Capacity}), at the expense of a higher local decoding complexity.
For example, for $R=0.667$ the distribution $\Lambda_9(x)$ (based on $k=3$) achieves $G^*(\mathcal C_9, \mbox{\boldmath $\Lambda$}_9)=0.465$, whereas 
$G^*(\mathcal C_{10}, \mbox{\boldmath $\Lambda$}_{10})=0.505$ is achieved by the distribution
$\Lambda_{10}(x)$ (based on $k=4$). In Fig.~\ref{fig:Capacity}
the thresholds achieved by regular schemes based on \ac{SPC} codes
of increasing rates are also shown. As $k$ grows the
rate of these scheme approaches $1$
and the corresponding threshold $1/(k+1)$ tends to $0$. For large $k$, the
scheme tends to operate close to the capacity bound for very high rates.


\section{Conclusions}\label{sec:Conclusions}

In this paper, a coding approach relying on iterative interference subtraction for the collision channel without feedback has been proposed and analyzed. The scheme, dubbed \ac{CSA}, can be seen as an extension of the \ac{IRSA} scheme, where the extension consists of splitting packets into segments and encoding the segments via randomly picked local component codes. A bridge between erasure decoding for graph-based codes and the iterative interference cancellation process of \ac{CSA} has been established, allowing an elegant analysis of the access scheme performance. Exploiting this graphical representation, density evolution equations for \ac{CSA} on the collision channel have been obtained and used to analyze the iterative interference subtraction process. The ``capacity'' of the \ac{CSA} scheme without retransmissions has been defined and, in the process, it has been shown that the scheme is asymptotically reliable even if retransmissions are forbidden. A throughput as high as $1\,\mathrm{[packets/slot]}$ has been shown to be tightly approachable when sufficiently low coding rates are employed for the component codes. Furthermore, a technique to design \ac{CSA} schemes with arbitrarily high coding rates has been developed which allows approaching the capacity bound over the whole range of rates. Numerical results have been presented to validate the proposed analytical framework.

We conclude this paper by discussing some possible directions of further investigation emerging from the presented results. Considering the same collision channel model adopted in the present paper, for example, the analogy with iterative decoding on the erasure channel suggests that it might be possible to develop sequences of \ac{CSA} configurations achieving the capacity bound \eqref{eq:capacity_bound} for any value of the rate $R$, similarly to the well-known \ac{LDPC} ``capacity-achieving sequences'' \cite{shokrollahi00:capacity-achieving}. We conjecture that, provided such sequences exist, their construction requires an increasing value of the component codes dimension $k$. Considering again the collision channel, the extra-ordinary performances obtained in the \ac{LDPC} coding context by exploiting spatial coupling \cite{SpatialCoupling:kudekar_TIT} prompt the adoption of this paradigm toward the design of ``convolutional'' \ac{CSA} schemes. (The only paper we are aware of in this context is \cite{Liva2012:ISIT} in which, however, only spatially coupled \ac{IRSA} configurations have been addressed.) Interesting directions of investigation also arise both from introducing spatial diversity through the assumption of availability of multiple receivers (as was done in \cite{Munari13:ALOHA_diversity} for pure slotted ALOHA) and from abandoning the simple collision channel model to consider the more general \ac{MPR} model. Some work in this sense has already been carried out in \cite{DeGaudenzi09:CRDSA}, in the framework of \ac{CRDSA} exploiting the capture effect, and in \cite{Ghanbarinejad13:IRSA_multiuser}, in the framework of \ac{IRSA} with multiuser detection.

\appendices

\section{Results on Successive Interference Cancellation}\label{sec:Appendix_IC}
In this appendix we address the actual performance achievable under a realistic \ac{SIC} scheme. 
More specifically, we intend to validate the assumption that, after removing $l-1$ interfering segments from a slice in which $l$ segments collided, 
the remaining segment can be decoded correctly with very high probability. In practice, this turns into verifying that the residual interference after 
interference cancellation (due to imperfect channel estimation) does not degrade considerably the performance of the error correcting code used to protect the segments.
 
To this purpose, let's consider the case where $l$ users attempt a segment transmission
within the same slice. We stick to the case of perfect power control and equal channel condition (gain) among the users. We denote by $u^{(i)}(t)$ the complex baseband \ac{PAM} signal transmitted by the
$i$-th user, i.e.,
\[
u^{(i)}(t)=\sum_{v=1}^{N_s}b^{(i)}_v
\gamma(t-v T_s)
\]
where $N_s$ is the number of symbols per
segment, $\{b^{(i)}_v\}$ is the sequence of such symbols and $T_s$ is
the symbol period. By
$\gamma(t)=\mathcal{F}^{-1}\left\{\sqrt{\mathtt{RC}(f)}\right\}$ we denote
the pulse shape, where $\mathtt{RC}(f)$ the frequency response of the
raised-cosine filter.

Each contribution is received with a random delay $\epsilon_i$, a
random frequency offset $f_i\sim \mathcal{U}[-f_{\max},f_{\max}]$ and
a random phase offset $\phi_i\sim
\mathcal{U}[0,2\pi)$. The received signal after
the \ac{MF} is given by $r(t)=\sum_{i=1}^l z^{(i)}(t)\ast h(t) +
n(t)$ where $n(t)$ is the Gaussian noise contribution,
$h(t)=\gamma^*(-t)$ is the \ac{MF} impulse response and
$z^{(i)}(t)=\sum_{v=1}^{N_s}b^{(i)}_v\gamma(t-vT_s-\epsilon_i)\exp(j2\pi
f_it + j\phi_i)$. Assuming frequency shifts that are small w.r.t.
the signal bandwidth (i.e., $f_{\max}T_s\ll 1$), the received signal
may be approximated by
\begin{equation}
r(t) \approx \sum_{i=1}^{l}\tilde{u}^{(i)}(t-\epsilon_i)e^{j2\pi f_i t
+ j\phi_i}+n(t)
\label{eq:RX_signal}
\end{equation}
where $\tilde{u}^{(i)}(t)$ is the response of the \ac{MF} to
${u}^{(i)}(t)$. In the following we regard $\tilde{u}^{(1)}(t)$ as the useful term and 
$\tilde{u}^{(2)}(t),\tilde{u}^{(3)}(t), \ldots,\tilde{u}^{(l)}(t)$
as the interference to be cancelled. These latter $l-1$ terms are assumed to have been 
successfully recovered via \ac{MAP} erasure decoding of the associated component code.


To proceed with \ac{SIC}, it is necessary to estimate the set of
parameters $\{\epsilon_i, f_i, \phi_i\}$, for $i \in \{2, \ldots, l\}$. As
suggested in \cite{DeGaudenzi07:CRDSA}, we consider the case where
$\epsilon_i$ and $f_i$ can be accurately estimated on the segments of the same
burst that have already been recovered, and
that their values remain constant through the frame. As pointed out
in \cite{DeGaudenzi07:CRDSA}, this argument does not hold for the
phase rotation terms $\phi_i$, which may not be stable from a slice
to another one. As such, we need to estimate $\phi_i$ for each
segment individually and directly on the slice where we want to eliminate its
contribution. A fine phase estimation can be obtained
by a data aided approach. Recall in fact that the symbol sequences
$\{b^{(i)}_v\}$ (for $i \in \{2\ldots l \}$) are known at the receiver, since
they can be reconstructed after \ac{MAP} erasure decoding of the associated component code. The \ac{SIC} works as follows. We denote by $y^{(i)}(t)$ the
signal at the input of the phase estimator for the $i$-th
contribution. In the first step, the input signal is given by
$y^{(2)}(t)=r(t)$ and the phase of the first interfering user
($i=2$) is estimated as
\[
\hat{\phi}_2=\arg \left\{\sum_{v=1}^{N_s}y^{(2)}_v\left(b_v^{(2)}\right)^*\right\}
\]
with
\[
y^{(2)}_v=y^{(2)}(vT_s+\epsilon_2)e^{-j2\pi f_2 (v T_s+\epsilon_2)}.
\]
After the estimation of the phase offset for the first interferer,
the corresponding signal can be reconstructed as
$\tilde{u}^{(2)}(t-\epsilon_2)e^{j2\pi f_2 t + j\hat{\phi}_2}$ and
its contribution can be removed from \eqref{eq:RX_signal}, i.e.
\[
y^{(3)}(t)=y^{(2)}(t)-\tilde{u}^{(2)}(t-\epsilon_2)e^{j2\pi f_2 t + j\hat{\phi}_2}.
\]
The \ac{SIC} proceeds serially. For the generic $i$-th contribution we have
\begin{equation}
\hat{\phi}_i=\arg \left\{\sum_{v=1}^{N_s}y^{(i)}_v\left(b_v^{(i)}\right)^*\right\}\label{eq:phiest}
\end{equation}
with $y^{(i)}_v=y^{(i)}(vT_s+\epsilon_i)\exp\left(-j2\pi f_i
(vT_s+\epsilon_i)\right)$ and
\[
y^{(i)}(t)=y^{(i-1)}(t)-\tilde{u}^{(i-1)}(t-\epsilon_{i-1})e^{j2\pi f_{i-1} t + j\hat{\phi}_{i-1}}.
\]
After the cancellation of the $l-1$  contributions the residual
signal, denoted by $y^{(1)}(t)$, is given by the $1$-st
user's contribution, the  noise $n(t)$, and a residual interference
term $\nu(t)$ due to the imperfect estimation of the interferers'
phases (causing imperfect \ac{SIC}), i.e.,
\begin{equation}
y^{(1)}(t)=\tilde{u}^{(1)}(t-\epsilon_1)e^{j2\pi f_1 t
+ j\phi_1}+n(t)+\nu(t).\label{eq:RX_final}
\end{equation}
The estimation of $\{\epsilon_1, f_1, \phi_1\}$ is then performed on
the signal in \eqref{eq:RX_final}. After sampling, soft-demodulation
takes place, and the log-likelihood ratios for the codeword bits are
derived. This data aided
approach works if the cross-correlation between the sequences
$\{b^{(i)}_v\}$, $i \in \{1\ldots l\}$, is on average low, which is
the case if each user encodes segments whose bits
can be modeled as \ac{i.i.d.} random variables.

We simulated the \ac{SIC} process  with various numbers
of collisions. The information sequences were randomly generated, then encoded through
a  $(512,256)$ cycle code from \cite{Liva13:Turbo} over $\mathbb F_{256}$. \ac{QPSK} modulation was
considered for the simulations. For each transmission attempt we generated the
parameters $\{\epsilon_i, f_i, \phi_i\}$ according to the
distributions presented before, with maximum frequency shift $f_{\max}=0.01/T_s$. The received signal $r(t)$ has then been
oversampled at a rate $M_s/T_s$ with $M_s=8$, and the
\ac{SIC} algorithm has been applied to the
oversampled digital signal.

\begin{figure}[t]
\begin{center}
\includegraphics[width=0.95\columnwidth,draft=false]{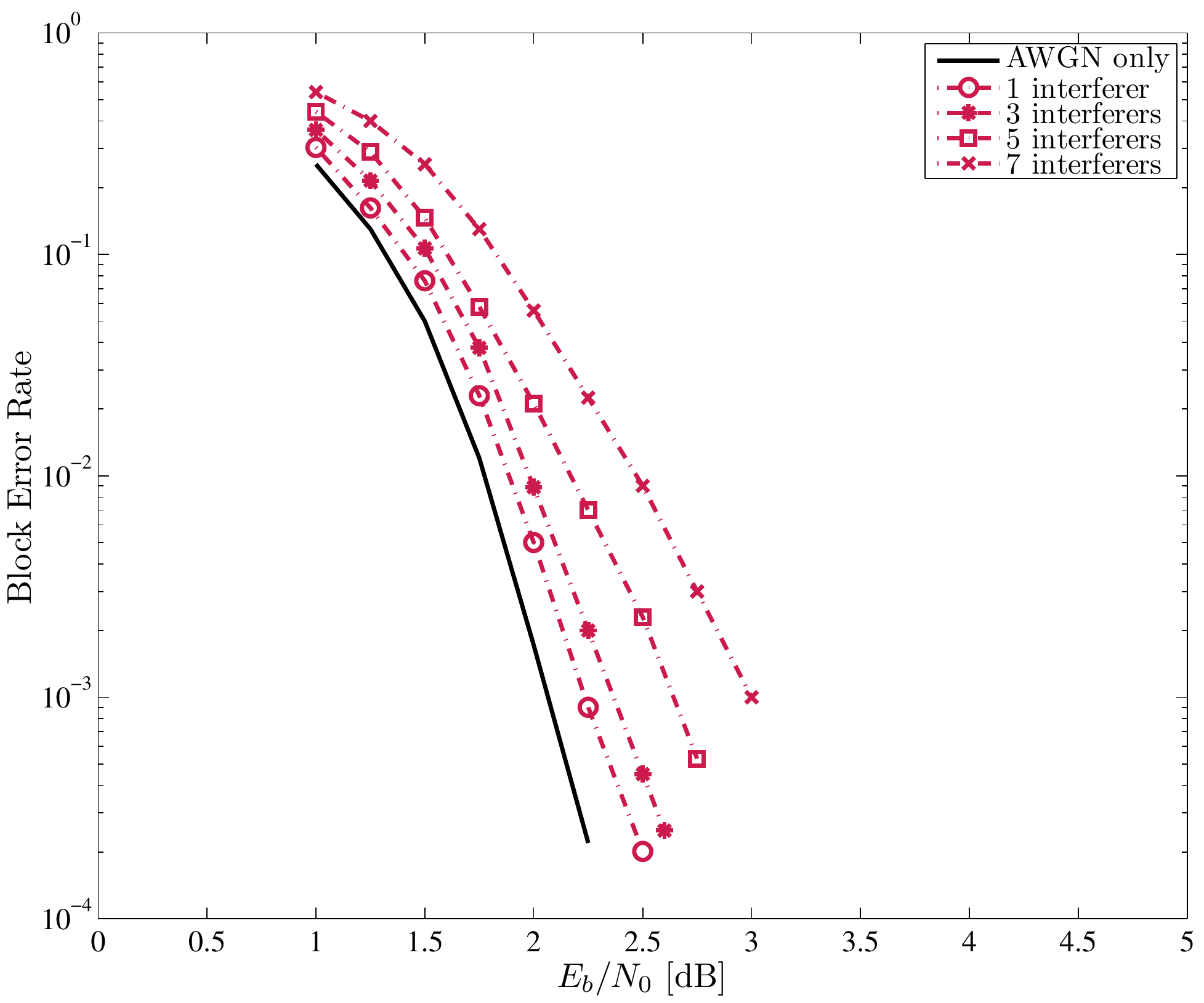}
\end{center}
\caption{Block error rate vs. $E_b/N_0$ for a $(512,256)$ cycle code with QPSK modulation, $50$ iterations of the belief propagation algorithm. Various number of interferers.}\label{fig:FER}
\end{figure}

Once the $l-1$ interference
contributions have been cancelled, log-likelihood ratios for the
codeword bits have been input to the channel decoder. In Fig.
\ref{fig:FER} the impact of the \ac{SIC} process on the block error rate for
the segment to be recovered (i.e., the signal corresponding to $i=1$) is shown in
terms of block error rate vs. $E_b/N_0$ for $l=2, 4, 6, 8$ segment collisions (i.e., $1,3,5,7$ interferers).
The performance on the \ac{AWGN} channel without collisions is
provided as reference. Note that, up to $l=8$ collisions, the
performance degradation due to the imperfect estimation of the phase
offsets is small, namely, less than $1$ dB at block error rate $\simeq
10^{-3}$. Considering $E_b/N_0=2.5$ dB, after removing
$l-1=7$ interference contributions we have a block error rate close to $10^{-2}$.

\section{An alternative Proof of the Capacity Bound \eqref{eq:capacity_bound}}\label{appendix:proof_2}

In this appendix, we propose an alternative proof of the upper bound \eqref{eq:capacity_bound}. This proof is based on adopting an equivalent channel model that is addressed next.

Encoded segments are packets of bits that can be mapped onto the elements of a 
finite field $\mathrm{GF}(2^l)$, for appropriate integer $l$. We model collisions between segments 
as sums of symbols in $\mathrm{GF}(2^l)$. In each slice of the \ac{MAC} frame
the decoder is capable to discriminate between a ``silence'' (no active user has transmitted in that segment), a symbol
in $\mathrm{GF}(2^l)$ corresponding to a unique slice, or a symbol in $\mathrm{GF}(2^l)$ 
being the result of a collision. In this latter case, the observed symbol 
in $\mathrm{GF}(2^l)$ provides no information to the decoder about the number 
and the values of colliding segments.

As such, with respect to the channel model discussed 
in Section~\ref{subsection:channel_model}, Assumption~\ref{assumption:1} and Assumption~\ref{assumption:2} remain valid, while Assumption~\ref{assumption:3}
is replaced by the following equivalent assumption (in that all developed results still hold):

\medskip
\begin{assumption} 
If a collision occurs between $d>1$ slices $s_1,s_2,\dots,s_d \in \mathrm{GF}(2^l)$, 
the symbol $s = s_1+s_2+\dots +s_d \in \mathrm{GF}(2^l)$ is generated in 
the corresponding segment of the frame. Cancellation of the interference contribution 
of a segment consists of adding the corresponding element 
of $\mathrm{GF}(2^l)$ to the current symbol in the associated slice of the frame.
\end{assumption}

\medskip
This channel model is then similar to an $F$-adder 
channel \cite{urbanke1993:Fadder}, with the difference that collisions may or may not occur
and that, when collisions take place, the decoder
can detect them.\footnote{It is worth pointing out that this simplified setting also represents a possible channel model for \emph{shared memories}, provided some mechanism is employed to discriminate between memory locations in which 
the data of a single users are stored and memory locations in which the data of several users are XORed.}

The upper bound \eqref{eq:capacity_bound} may now be derived as a simple consequence of the Rouch{\'e}-Capelli Theorem. 
Regarding the information segments of the active users as the unknowns of a linear system of equations\footnote{Recall, in fact, that each encoded segment may be expressed as a linear combination of the associated information segments.} and the symbol in $\mathrm{GF}(2^l)$ available in a non-empty slice as the known term of the corresponding equation, the system admits no unique solution whenever the number of unknowns exceeds the number of available equations. As $M \rightarrow \infty$ the expected fraction of non-empty slices is $1-\Psi_0=1-\exp \{ -G/R \}$, while the expected number of unknowns per slice is equal to the expected channel load $G$ which yields
$$
G \leq 1 -e^{ -G/R }
$$
as a necessary condition for successful decoding. This inequality is equivalent to \eqref{eq:R_bound}, the proof remaining the same hereafter. 

It is pointed out that a similar proof technique was adopted in \cite[Section~II-D]{Popovski2012:comlet} to upper bound the success probability of the frameless scheme there considered. As it was recognized in \cite{Popovski2012:comlet}, the bound there obtained represents a special instance of the capacity bound presented in this paper.

\section*{Acknowledgment}
The authors would like to thank the Anonymous Reviewers and the Associate Editor for their insightful technical comments which helped to improve the paper.

\end{document}